\documentclass[12pt]{article}
\pdfoutput=1
\usepackage{graphicx,psfrag,epsf,color}
\usepackage{amsmath,amssymb,amsfonts}
\usepackage{array}
\usepackage{cite}
\usepackage{rotating}
\usepackage{slashed, cancel}
\usepackage{xcolor}
\usepackage[caption=false]{subfig}
\usepackage{graphicx}
\usepackage{mathrsfs}  
\usepackage{tabularx}
\newcolumntype{C}{>{\centering\arraybackslash}X}
\usepackage{multirow} \usepackage{array}

\numberwithin{equation}{section}

\newcounter{MBQ}

\newcommand{\nm}{n_-}

\newcommand{\eps}{\epsilon}

\def\be{\begin{equation}}
\def\ee{\end{equation}}
\def\beq{\begin{eqnarray}}
\def\eeq{\end{eqnarray}}
\newcommand{\bea}{\begin{eqnarray}}
\newcommand{\eea}{\end{eqnarray}}
\newcommand{\beas}{\begin{eqnarray*}}
\newcommand{\eeas}{\end{eqnarray*}}

\newcommand{\alem}{\alpha_{\rm em}}

\newcommand{\bra}[1]{\big\langle{#1}\big\vert}
\newcommand{\ket}[1]{\big\vert{#1}\big\rangle}

\DeclareSymbolFont{matha}{OML}{txmi}{m}{it}
\DeclareMathSymbol{\varv}{\mathord}{matha}{118}

\DeclareMathOperator{\T}{T}  

\newcommand{\lep}{\mathrm{\ell}} 
\newcommand{\bv}{\bar{\nu}} 

\newcommand{\Fa}{F_{A_\perp}}
\newcommand{\Fv}{F_{V}}
\newcommand{\Fn}{F_{A_\parallel}}
\newcommand{\hatFa}{\hat{F}_{A_\perp}}
\newcommand{\hatFn}{\hat{F}_{A_\parallel}}
\newcommand{\nplusq}{n_+q}
\newcommand{\nminusq}{n_-q}
\newcommand{\lambdaBp}{\lambda_B^+(n_-q)}
\newcommand{\lambdaBm}{\lambda_B^-(n_-q)}

\newcommand{\ga}{\gamma}

\newcommand{\LQCD}{\Lambda_{\text{QCD}}}

\usepackage{floatrow}
\DeclareFloatFont{small}{\footnotesize}

\begin{document}
\allowdisplaybreaks

\begin{titlepage}

\begin{flushright}
{\small
TUM-HEP-1317/21\\
Nikhef-2021-009\\
February 19, 2021\\
}
\end{flushright}

\vskip1cm
\begin{center}
{\Large \bf\boldmath QCD factorization of the 
four-lepton decay\\ $B^-\rightarrow \ell \bar{\nu}_\ell 
\ell^{(\prime)} \bar{\ell}^{(\prime)}$}
\end{center}

\vspace{0.5cm}
\begin{center}
{\sc Martin~Beneke,$^a$ Philipp B\"oer,$^a$ 
Panagiotis Rigatos,$^a$ K. Keri Vos$^{b,c}$} \\[6mm]
{\it $^a$Physik Department T31,\\
James-Franck-Stra\ss{}e~1, 
Technische Universit\"at M\"unchen,\\
D--85748 Garching, Germany} \\[0.3cm]

{\it $^b$Gravitational 
Waves and Fundamental Physics (GWFP),\\ 
Maastricht University, Duboisdomein 30,\\ 
NL-6229 GT Maastricht, the
Netherlands}\\[0.3cm]

{\it $^c$Nikhef, Science Park 105,\\ 
NL-1098 XG Amsterdam, the Netherlands}
\end{center}

\vspace{0.6cm}
\begin{abstract}
\vskip0.2cm\noindent
Motivated by the first search for the rare charged-current $B$ 
decay to four leptons, 
$\ell \bar{\nu}_\ell \ell^{(\prime)} \bar{\ell}^{(\prime)}$, 
we calculate the decay amplitude with factorization methods. 
We obtain the $B\to \gamma^*$ form factors, which depend on the 
invariant masses of the two lepton pairs, at leading power in an 
expansion in $\Lambda_{\rm QCD}/m_b$ to next-to-leading order 
in $\alpha_s$, and at $\mathcal{O}(\alpha_s^0)$ 
at next-to-leading power. Our calculations predict branching 
fractions of a few times $10^{-8}$ in the $\ell^{(\prime)} \bar{\ell}^{(\prime)}$ mass-squared bin up to $q^2=1~$GeV$^2$ with $n_+q>3~$GeV. 
The branching fraction rapidly 
drops with increasing $q^2$. An important 
further motivation for this investigation has been to explore 
the sensitivity of the decay rate to the inverse moment  
$\lambda_B$ of the leading-twist $B$ meson light-cone 
distribution amplitude. We find that in the small-$q^2$ bin, 
the sensitivity to $\lambda_B$ is almost comparable to 
$B^- \rightarrow \lep^- \bv_{\lep}\gamma$ when $\lambda_B$ is small, 
but with an added 
uncertainty from the light-meson intermediate resonance 
contribution. The sensitivity degrades with larger 
$q^2$.
\end{abstract}
\end{titlepage}


\section{Introduction}

The radiative decay $B^- \rightarrow \lep^- \bv_{\lep}\gamma $ has 
been extensively studied in the context of QCD factorization (QCDF) 
\cite{Korchemsky:1999qb, DescotesGenon:2002mw,Lunghi:2002ju, Bosch:2003fc,Beneke:2011nf} when the energy of the photon $E_\gamma$ 
is large compared 
to the scale of the strong interaction $\Lambda_{\text{QCD}}$. At 
leading power in a simultaneous expansion in $\LQCD/E_{\gamma}$ 
and $\Lambda_{\text{QCD}}/m_b$, and at leading order (LO) in 
the strong coupling $\alpha_s$, the relevant $B\to \gamma$ 
transition form factor can be expressed in terms of only two 
hadronic parameters: the accurately known $B$ meson decay constant 
$f_B$, and the poorly constrained first inverse moment  
$1/\lambda_B = \int_0^{\infty} d\omega \,\phi^B_+(\omega)/\omega$ 
of  $\phi^B_+(\omega)$, the leading-twist $B$ meson light-cone 
distribution amplitude 
(LCDA). This hadronic parameter was introduced 
in the theoretical description of charmless hadronic decays 
\cite{Beneke:1999br} and appears in the QCD calculation of almost 
any other exclusive $B$ decay to light particles. The radiative decay 
$B^- \rightarrow \lep^- \bv_{\lep}\gamma $ has been advocated 
as a means to determine $\lambda_B$ from data 
\cite{Beneke:2011nf}. First significant measurements can be 
expected from the BELLE~II experiment (see \cite{Gelb:2018end}
for the most recent BELLE result).

This strategy is difficult to implement in the hadronic 
$B$ experiment LHCb, since the photon in the radiative decay 
cannot be easily reconstructed. In this paper we investigate 
whether the four-lepton decay $B^-\rightarrow 
\lep \bv_{\lep} \gamma^* \rightarrow
\ell \bar{\nu}_\ell \ell^{(\prime)} \bar{\ell}^{(\prime)}$, 
in which the real photon is replaced by a virtual one, 
which  decays into a lepton-antilepton pair ($\lep,\lep'= e,\mu$), 
retains sensitivity to $\lambda_B$, and hence could 
provide an alternative measurement. We focus on the kinematic region, 
where the $\gamma^*$, respectively the lepton pair, has large energy 
but small invariant mass $q^2\lesssim 6\,\mbox{GeV}^2$.\footnote{In case of identical 
lepton flavours $\ell=\ell^\prime$, the identification 
of the virtual photon with a lepton-antilepton pair is 
not unique, see later discussion.} The four-lepton decays 
have not been observed up to now, but the LHCb 
experiment~\cite{Aaij:2018pka} established an upper 
bound of $\mbox{Br}\,(B^+ \rightarrow \mu^+ \bv_{\mu} \mu^- \mu^+) \, 
< 1.6 \cdot 10^{-8} $ on the branching fraction of the muonic 
mode under the assumption that the 
smaller of the two possible $\mu^+ \mu^-$ invariant masses 
is below 980 MeV, which is close to, in fact somewhat below, 
theoretical expectations \cite{Danilina:2018uzr, Danilina:2019dji}.

The factorization theorem for the $B \to \gamma$ form 
factors in the regime where the photon is energetic, 
$\nplusq =2 E_\gamma 
\gg \Lambda_{\rm QCD}$, has been established long ago 
\cite{Lunghi:2002ju, Bosch:2003fc}. Its generalization 
to $B \to \gamma^*$ form factors is straightforward, 
when $q^2$ is away from light-meson 
resonances. The present treatment follows the strategy 
applied to  $B^- \rightarrow \lep^- \bv_{\lep}\gamma $ 
\cite{Beneke:2011nf} and $B_s\to \mu^+\mu^-\gamma$ 
\cite{Beneke:2020fot} -- we compute the form factor 
in QCD factorization at leading power (LP) including 
$\mathcal{O}(\alpha_s)$ QCD corrections, and include 
next-to-leading power (NLP) corrections at
 $\mathcal{O}(\alpha_s^0)$. 
The light-meson resonance contribution is 
included in the same fashion as for the ``type-B'' contribution 
to $B_s\to \mu^+\mu^-\gamma$ \cite{Beneke:2020fot}. 
Since the four-lepton final state is produced from a virtual 
$W$ boson {\em and} photon, an extension of previous calculations 
is required to $B \to \gamma^*$ form factors that 
depend on two non-vanishing virtualities. We note that a 
previous computation \cite{Albrecht:2019zul} of  
these $B \to \gamma^*$ form factors includes either {\em only} 
the resonance contribution at small $q^2$, or employs QCD sum rules 
that apply only to large $q^2\sim m_b^2$. No attempts have 
so far been undertaken to estimate the form factors 
for intermediate and small $q^2$ with factorization methods, 
as done here, which apply when 
$\nplusq \gg \Lambda_{\rm QCD}$. With these kinematic 
restrictions the differential branching fraction of 
the four-lepton decay is expressed, at LP, in terms of 
generalized inverse moments of the $B$ meson LCDA, 
which can be related to~$\lambda_B$.

We consider the case of non-identical lepton flavours, 
$\ell^\prime\not=\ell$, and identical ones, which requires 
additional kinematic considerations.


\section{Basic definitions}

Following the conventions of \cite{Beneke:2011nf}, we write the 
$B^- \to \ell \bar{\nu}_\ell  \ell' \bar{\ell}'$ decay amplitude 
to lowest non-vanishing order in the electromagnetic coupling as
\begin{align}\label{eq:ampl}
\mathcal{A}\left(B^- \to \ell \bar{\nu}_\ell  \ell' \bar{\ell}'\right)&=\frac{G_F V_{u b}}{\sqrt{2}}\bra{\ell (p_\ell) \, \bar{\nu}_\ell (p_\nu) \, \ell' (q_1)\,  \bar{\ell}'(q_2)}\bar{\ell} \gamma^{\mu}\left(1-\gamma_{5}\right) \nu_{\ell} \cdot \bar{u} \gamma_{\mu}\left(1-\gamma_{5}\right) b \ket{B^{-}(p)}  \nonumber \\
&= \frac{G_F V_{u b}}{\sqrt{2}}\frac{ie^2}{q^2}\,Q_{\ell'}\left[T^{\mu \nu}(p,q)+Q_\ell f_B\, g^{\mu \nu} \right] \left( \bar{u}_{\ell'} \ga_{\mu} v_{\bar{\ell'}} \right)  \left( \bar{u}_{\ell} \ga_{\nu}(1-\gamma_5) v_{\bar{\nu}} \right) \ , 
\end{align}
where $q\equiv q_1+q_2$ and $p=m_B v = q + k$, such that  
$k = p_\ell + p_\nu$ is the momentum of the virtual $W$ boson. 
In addition, we use the convention 
$iD^\mu = i \partial^\mu - Q_\ell e A_{\rm em}^\mu$ for the 
electromagnetic covariant derivative, with $Q_\ell=-1$ for 
the lepton fields. The hadronic tensor
\begin{equation}
T^{\mu \nu}(p,q) = \int d^4 x\, e^{i q \cdot x} \bra{0} \T \left\{j^{\mu}_{ \rm em}(x) \left(\bar{u}\ga^{\nu} \left(1-\ga^5\right)b\right)(0) \right\} \ket{B^-} \ , 
\label{eq:hadtensor}
\end{equation}
with the electromagnetic current $j^\mu_{\rm em} = \sum_q Q_q \bar{q}\gamma^\mu q + Q_\ell \bar{\ell}\gamma^\mu \ell$ accounts for the 
emission of the virtual photon from the $B$ meson constituents. 
The second term in the square brackets in \eqref{eq:ampl} 
corresponds to the emission from the final-state lepton, 
see Fig.~\ref{fig:diagrams} below. It can be 
expressed in terms of the $B$ meson decay constant
$\bra{0}\bar{u}\gamma^{\nu} \left(1-\gamma^5\right)b \ket{B^-(p)} = -i f_B p^{\nu}$ 
and constitutes a power correction relative to the $T^{\mu\nu}$ 
term in the kinematic region of interest. 

The hadronic tensor $T^{\mu\nu}$ can be decomposed into six form 
factors $F_i(q^2,k^2)$ of two kinematic invariants. Applying the 
Ward identity $q_{\mu}T^{\mu \nu}=  f_B p^{\nu}$ leaves four form 
factors and a contact term (see App.~\ref{sec:app:Tdecomp} for 
details). We write 
\begin{align}
T^{\mu \nu}(p,q) &= 
\left(g^{\mu \nu }v\cdot q - v^{\mu} q^ {\nu}\right) \hatFa 
+i\,  \epsilon^{\mu \nu \alpha \beta}\, v_{\alpha} q_{\beta} \Fv
-\hatFn v^\mu q^\nu    + \mbox{$(q^\mu, k^\nu)$ terms}\,.
\label{eq:TintoFF}
\end{align}
We neglect the lepton masses, in which case the $q^\mu, k^\nu$ 
terms drop out after contracting $T^{\mu \nu}(p,q)$ with the lepton 
tensor. The contact term is fixed by the Ward identity to 
$(f_B m_B)/(v\cdot q) v^\mu v^\nu$. This can be rewritten as $f_B/(v\cdot q) \,v^\mu (k^\nu+q^\nu)$ and has been absorbed into 
$\hat{F}_{A_\parallel}$ and the $k^\nu$ terms in (\ref{eq:TintoFF}). 
The convention for the totally anti-symmetric tensor is 
$\epsilon^{0123}=1$. The virtual photon emission from the 
final-state lepton $\ell$ in \eqref{eq:ampl} 
is exactly cancelled by the redefinition
\begin{equation}
    \Fa=\hatFa + \frac{Q_\ell f_B}{v \cdot q} \ , \quad\quad \tilde{F}_{A_\parallel}=\hatFn- \frac{Q_\ell f_B}{v \cdot q} \, .
\end{equation}
Therefore, the term in square brackets 
in \eqref{eq:ampl} can be expressed in terms of three form factors. 
To separate amplitudes corresponding to the different polarization 
states of the virtual photon, we shall use the decomposition 
\begin{equation}\label{eq:newpara}
T^{\mu \nu}(p,q)+Q_\ell f_B g^{\mu\nu}= \Fa \, g^{\mu \nu }_{\perp} v\cdot q  +i\, \Fv \epsilon^{\mu \nu \alpha \beta}\, v_{\alpha} q_{\beta} - \Fn \,v^{\mu} q^{\nu}  \,,
\end{equation}
which implies
\begin{equation}
\label{eq:redefff}
  \Fn =  \tilde{F}_{A_{\parallel}}- \frac{2q^2(m_B^2-q^2+k^2)}{\lambda}\Fa \,.
\end{equation}
 Here $\lambda \equiv \lambda(m_B^2,q^2,k^2) = m_B^4 - 2m_B^2 (k^2 + q^2) + (k^2-q^2)^2$ the K\"all\'en function. The form factor $\Fn$ arises from a longitudinally polarized virtual photon and vanishes in the real-photon limit $q^2 \to 0$. 
Without loss of generality we choose 
the three-momentum $\vec{q}$ to point in the positive $z$ direction, 
such that its decomposition into light-cone vectors $n_\pm^\mu$ reads
\begin{align}
q^\mu = \nplusq \,\frac{n_-^\mu}{2} + \nminusq \,\frac{n_+^\mu}{2} 
\ , 
\end{align}
with $n_\pm^\mu = (1,0,0,\mp 1)$ and $q^2 = \nplusq \,\nminusq$. 
The transverse metric tensor is then $g^{\mu \nu}_{\perp}=g^{\mu \nu}- (n_+^{\mu}n_-^{\nu}+n_+^{\nu}n_-^{\mu})/2$. The large component 
$\nplusq$ of $q^\mu$ is related to the invariant masses $q^2$ 
and $k^2$ via
\begin{equation}
\label{eq:qplusk2}
\nplusq = \frac{m_B^2  - k^2 + q^2 + \sqrt{\lambda}}{2m_B} \ .
\end{equation}
Finally, we define the left- and right-helicity form factors
\begin{equation}\label{eq:lrffs}
F_{L} = \frac{1}{2} \left(  \Fv + \Fa \right), 
\quad\quad     
F_{R}= \frac{1}{2} \left(  \Fv - \Fa \right). 
\end{equation}
Since helicity is conserved in high-energy QCD processes, 
$F_R$ is power-suppressed relative to $F_L$ in the heavy-quark / 
large $\nplusq$ limit.

For non-identical lepton flavours $\ell\neq \ell'$, the differential decay 
width can be obtained analytically by a straightforward calculation.
The full angular distribution, i.e. the five-fold differential rate 
is given in App.~\ref{app:angdist}. Here we quote the 
double differential rate in the invariant masses $q^2$ and $k^2$,
for which we obtain the simple expression
\begin{align}
\label{eq:d2Br}
\frac{d^2 {\rm Br}\left(B^{-} \rightarrow \ell \, \bar{\nu}_\ell 
\, \ell' \bar{\ell'}\right)}{dq^2\,dk^2} 
&=\frac{\tau_BG_F^2|V_{ub}|^2\alem^2}{2^{8}3^2\pi^3m_B^5}
\frac{\sqrt{\lambda}}{q^2} \,\sqrt{1-\frac{4m_{\lep'}^2}{q^2}}
\left(1-\frac{m_{\lep}^2}{k^2}\right) \nonumber \\
\,&\times \left(8 k^2\left(m_B^2+q^2-k^2\right)^2
\left|\Fa\right|^2 
+8 k^2\lambda\left|\Fv\right|^2  
+ \frac{\lambda^2}{q^2} \,|F_{A_{\parallel}}|^2 \right)\,,
\end{align}
keeping the lepton masses $m_{\ell^{(\prime)}}$ in the phase space 
integration (implying $q^2 > 4 m_{\lep'}^2$), which is relevant 
for muons.

\begin{figure}[t]
\includegraphics[scale=0.8]{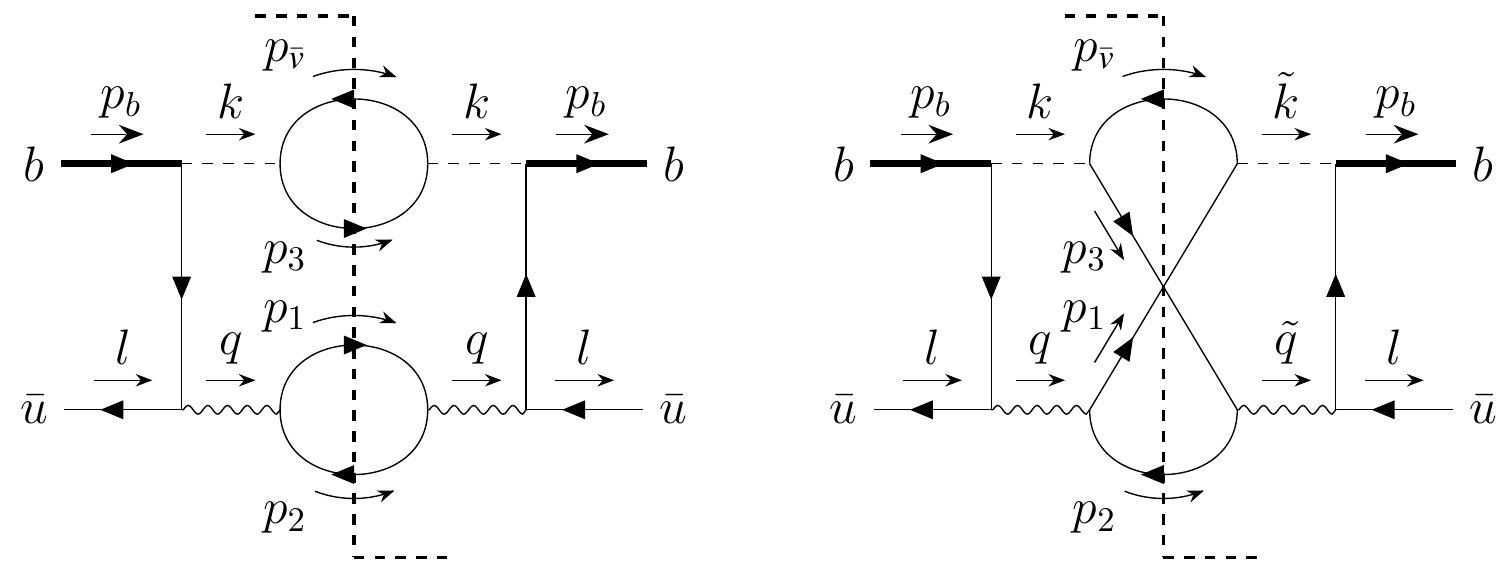}
\centering
\vskip0.2cm
\caption{Graphical representation of the squared amplitude for 
the non-identical-lepton final-state (left) and of the interference 
term for the case of identical lepton flavours (right).}
\label{fig:squareddiagrams}
\end{figure}

The case of identical lepton flavours $\ell^\prime=\ell$ is more
 complicated, as an additional contribution from the interchange of the two final-state leptons arises. To clarify this point, let 
us define $B^-\to \ell^-(p_1) \ell^+(p_2)\ell^-(p_3) \bar{\nu}_\ell(p_\nu)$, with $q^2=(p_1+p_2)^2$ and $\tilde{q}^2=(p_2+p_3)^2$. 
At the amplitude level, we have 
$\mathcal{M}_{\rm tot} = \mathcal{M}_a - \mathcal{M}_b$ 
where $\mathcal{M}_b = \mathcal{M}_a(p_1 \to p_3, p_3 \to p_1)$.
For the decay rate, this results in an additional interference 
term between $\mathcal{M}_a$ and $\mathcal{M}_b$, while the rates 
$\Gamma_{a,b} \propto |\mathcal{M}_{a,b}|^2$ from the squares of 
the individual diagrams are equal (as depicted in Fig.~\ref{fig:squareddiagrams}). Since $\Gamma_a + \Gamma_b$ is 
equal to the rate for non-identical lepton flavours, we find 
\begin{equation}\label{eq:brident}
    {\rm Br}\left(B^{-} \rightarrow \ell \, \bar{\nu}_\ell \, \ell \bar{\ell}\right) = {\rm Br}\left(B^{-} \rightarrow \ell \, \bar{\nu}_\ell \, \ell' \bar{\ell'}\right) +{\rm Br}_{\rm int}\left(B^{-} \rightarrow \ell \, \bar{\nu}_\ell \, \ell \bar{\ell}\right).
\end{equation}
For the interference term 
$d^2 {\rm Br}_{\rm int}\left(B^{-} \rightarrow \ell \, \bar{\nu}_\ell 
\, \ell \bar{\ell}\right)/(dq^2\,dk^2)$ can only be obtained 
numerically.

\section{Calculation of the form factors}

The amplitude can be factorized through an expansion in 
$\Lambda_{\rm QCD}/m_b$ and $\Lambda_{\rm QCD}/\nplusq$, if the 
quark propagator that connects the electromagnetic and the weak 
current is far off-shell. This happens for very large $q^2$ of order 
$m_B^2$ in which case the amplitude can be reduced to a hard 
matching coefficient times the $B$ meson decay constant defined as a 
local matrix element in heavy-quark effective theory (HQET). 
The decay rate 
for such large $q^2$ is highly suppressed. The situation is more 
interesting when $q^2 \ll m_B^2$, but $q^\mu$ has still a large 
component $\nplusq \sim \mathcal{O}(m_B)$, while $\nminusq \sim 
\mathcal{O}(\Lambda_{\rm QCD})$, or even smaller. In this case the 
intermediate quark propagator becomes hard-collinear and the 
$\gamma^*$ probes the light-cone structure of the $B$ meson.
A factorization formula, which expresses the form factors as a 
convolution of the $B$ meson LCDA with a perturbative 
scattering kernel, can 
then be derived for the LP contribution using soft-collinear 
effective theory \cite{Bauer:2000yr,Bauer:2001yt,Beneke:2002ph,Beneke:2002ni} 
by matching \mbox{QCD $\to$ SCET$_{\rm I}$ $\to$ HQET}. Since the 
derivation is very similar to the one for 
$B^- \rightarrow \lep^- \bv_{\lep}\gamma $ and 
$B_s\to \mu^+\mu^-\gamma$ 
decays~\cite{Beneke:2011nf, Beneke:2020fot}, we only 
sketch the main steps in the following.

Upon integrating out the hard scales $m_b, \nplusq$, the 
flavour-changing weak current is represented in SCET$_{\rm I}$ by 
\begin{align}
\label{eq:hardmatching}
 \bar{q} \gamma^\mu (1- \gamma_5) b \to C_V^{(A0)} \, [\bar{q}_{\rm hc} \gamma_\perp^\mu (1-\gamma_5) h_v]
 + \left( C_4 n_-^\mu + C_5 v^\mu \right) [\bar{q}_{\rm hc} (1+\gamma_5) h_v] \,,
\end{align}
with hard matching coefficients $C_i = C_i(\nplusq; \mu)$.
Here $q_{\rm hc} = W^\dagger \xi_{\rm hc}$ is the hard-collinear 
quark field in SCET, multiplied with a hard-collinear Wilson line 
to ensure SCET collinear gauge-invariance.
Fields without arguments live at space-time point $x = 0$.
At LP, the index $\mu$ is transverse, since the LP SCET$_{\rm I}$ 
electromagnetic current $j^\mu_{q, {\rm SCET}_{\rm I}}(x)$ \cite{Lunghi:2002ju} 
contains only the transverse polarization of the virtual photon.  
We therefore only need $C_V^{(A0)}$, which to $\mathcal{O}(\alpha_s)$ reads \cite{Bauer:2000yr}
\begin{eqnarray}
\label{eq:cv}
C_V^{(A0)}(\nplusq; \mu) &=& 
1 + \frac{\alpha_sC_F}{4\pi} \bigg( 
- 2 \ln^2 \frac{m_B z}{\mu} + 5 \ln \frac{m_B z}{\mu} - \frac{3 - 2z}{1-z} \ln z
\nonumber\\
&& -\,2 \,{\rm Li}_2(1-z) - \frac{\pi^2}{12} - 6 \bigg)\,,
\end{eqnarray}
with $\alpha_s\equiv \alpha_s(\mu)$ in the $\overline{\rm MS}$ scheme, and $z = \nplusq/m_B = 1 - k^2/m_B^2 + \mathcal{O}(\Lambda_{\rm QCD}/m_B)$. The hadronic tensor is then expressed as 
\begin{align}
\label{eq:TfulltoSCET1}
    T^{\mu\nu}(p,q) = 2 C_V^{(A0)} \, {\cal T}^{\mu\nu}(q) 
\end{align}
in terms of the matching coefficient and the SCET$_{\rm I}$ correlation function
\begin{align}
\label{eq:SCET1corrfcts}
 {\cal T}^{\mu\nu}(q) &= \int d^4 x \, e^{i q x} \bra{0} T \left\{ j^\mu_{q, {\rm SCET}_{\rm I}}(x), [\bar{q}_{\rm hc} \gamma_\perp^\nu P_L h_v](0) \right\} \ket{B^-_v} \,.
\end{align}
A discussion of the precise power counting of the individual terms 
in $j^\mu_{q, {\rm SCET}_{\rm I}}$ and the possibility of 
extrapolating the above expressions to the large $q^2$ region 
with tree-level accuracy can be found in~\cite{Beneke:2020fot}. 

The SCET$_{\rm I}$ correlation function is then matched at LP 
to HQET. This results in 
\begin{align}
\label{eq:Tfac}
 {\cal T}^{\mu\nu}(q) = \left( g_\perp^{\mu\nu} + \frac{i}{2}  
\eps^{\mu\nu\rho\sigma} n_{+ \rho} n_{- \sigma}\right) \frac{Q_u F_B m_B}{4} \, \int_0^\infty d \omega \, \phi^B_+(\omega) \, \frac{J(\nplusq, q^2, \omega; \mu)}{\omega - \nminusq - i 0^+}, 
\end{align}
where $\nminusq = q^2/\nplusq$. The hard-collinear matching function \cite{Wang:2016qii}
\begin{align}\label{eq:jfun}
 J(\nplusq, q^2,\omega;\mu) &= 1 + \frac{\alpha_s C_F}{4\pi} \,\Bigg\{ \ln^2 \frac{\mu^2}{\nplusq \,(\omega - \nminusq)} - \frac{\pi^2}{6} -1  \nonumber \\
 - \frac{\nminusq}{\omega} &\ln \frac{\nminusq-\omega}{\nminusq} \left[ \ln \frac{\mu^2}{-q^2} + \ln \frac{\mu^2}{\nplusq \,(\omega - \nminusq)} +3 \right] \Bigg\} 
\end{align}
is convoluted with the leading-twist $B$ meson LCDA 
$\phi^B_+(\omega)$ defined through 
\begin{align}
\label{eq:BLCDA}
 \bra{0} \bar{q}_s(t n_-) [t n_-, 0] \slashed{n}_- \gamma_5 h_v(0) \ket{B^-_v} = i m_B F_B \int_0^\infty d \omega \, e^{-i \omega t} \phi^B_+(\omega) \,,
\end{align}
which contains the scale-dependent HQET $B$ meson decay 
constant $F_B = F_B(\mu)$. Similar to the denominator 
in~\eqref{eq:Tfac}, $\nminusq$ in (\ref{eq:jfun}) must be 
understood as $\nminusq + i 0^+$.

As a consequence of helicity conservation of QCD in the high-energy limit, the Lorentz structure in~\eqref{eq:Tfac} gives a non-vanishing contribution only to the left-helicity form factor $F_L$, which can be expressed as
\begin{align}
\label{eq:lpFF}
 F_L^{\rm LP} = C_V^{(A0)}(\mu) \,
\frac{Q_u F_B(\mu) m_B}{\nplusq} \, \int_0^\infty d \omega \, \phi^B_+(\omega; \mu) \, \frac{J(\nplusq, q^2, \omega; \mu)}{\omega - \nminusq - i 0^+} \,.
\end{align}
The form factors $F_R$ and $F_{A_\parallel}$ vanish at leading power, $F_R^{\rm LP} = F_{A_\parallel}^{\rm LP} = 0$. The $ - i 0^+$ 
prescription in \eqref{eq:lpFF} generates a rescattering phase of 
the form factor $F_L$ for $q^2>0$.

\begin{figure}[t]
\includegraphics[scale=0.85]{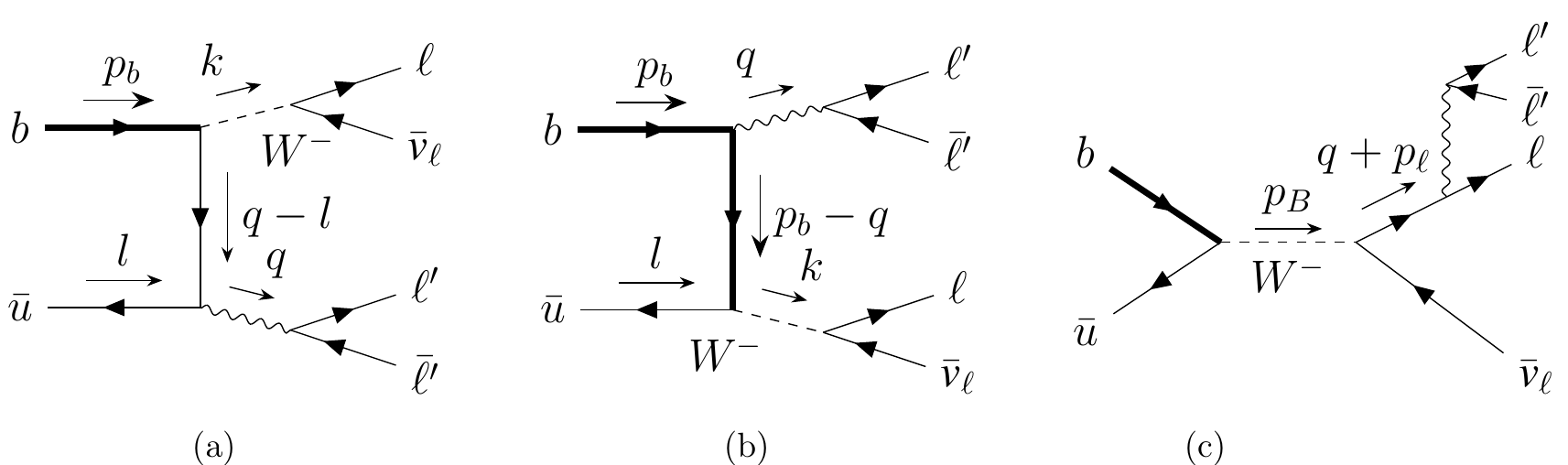}
\centering
\caption{Photon emissions that contribute to the tree-level $B^- \to \ell \bar{\nu}_\ell \ell' \bar{\ell}'$ amplitude.
The emission from the spectator quark (a) contributes at leading power whereas the emission from the heavy quark (b) and the lepton (c) is power suppressed.}
\label{fig:diagrams}
\end{figure}

A factorization formula for NLP corrections is presently not known. 
Following \cite{Beneke:2011nf, Beneke:2020fot} we infer  
the leading NLP $\mathcal{O}(\alpha_s^0)$ contributions by a 
diagrammatic analysis of the tree diagrams of Fig.~\ref{fig:diagrams}.
In the hard-collinear region, diagrams (b) and (c) vanish at 
LP. Their NLP contribution can be expressed in terms of $f_B$. 
In diagram (a), which gives the  $\mathcal{O}(\alpha_s^0)$ term in 
the above LP factorization result, we now expand the light-quark 
propagator to NLP, and obtain 
\begin{align}
\label{eq:propexpansion}
 \frac{(\slashed{q}-\slashed{l})}{(q-l)^2} = 
\frac{1}{\nminusq-\omega} \frac{\slashed{n}_-}{2}
 + \Bigg[
 \frac{1}{\nplusq}\frac{\slashed{n}_+}{2} - \frac{ \slashed{l}_\perp}{\nplusq (\nminusq - \omega)} + \frac{n_+l}{\nplusq} \frac{\omega}{(\nminusq - \omega)^2} \frac{\slashed{n}_-}{2}
 \Bigg] +\ldots\,,
\end{align}
where $l$ denotes the spectator-quark momentum of order 
$\Lambda_{\rm QCD}$ and terms suppressed by two powers of 
$\Lambda_{\rm QCD}/\{\nplusq,m_b\}$ are neglected. 
This expression reduces to the one \cite{Beneke:2011nf} for 
real photons, when $\nminusq = 0$ and $\nplusq = 2 E_\gamma$. 
Proceeding as in \cite{Beneke:2011nf}, we find 
\begin{align}
\label{eq:interpolL}
F_L^{\rm NLP}&= \xi(q^2,v\cdot q) + \frac{Q_{\ell}f_B}{2v \cdot q}\,, 
\\
\label{eq:interpolR}
F_R^{\rm NLP}&= \frac{F_B}{\nplusq}\frac{m_B Q_u}{\nplusq} 
\left(1+ \frac{\nminusq}{\lambdaBp }\right)
- \frac{F_B m_B Q_b}{q^2-2 m_b v \cdot q} 
- \frac{Q_{\lep}f_B}{2v \cdot q} 
\end{align}
for the $\mathcal{O}(\alpha_s^0)$ NLP terms of the form factors. 
These expressions are written in a form such that the complete 
expression for $F_{L,R}$ including $F_L^{\rm LP}$ is valid for 
both, hard-collinear and hard $q^2$.\footnote{
The absence of a soft form factor in the hard region is respected 
by our ansatz for $\xi(q^2,v\cdot q)$ below, which makes it a 
next-to-next-to-leading power correction in the hard region.} 
Setting 
$2v \cdot q =\nplusq+\nminusq \to\nplusq$ and neglecting 
$q^2$ in the denominator of the second term in 
(\ref{eq:interpolR}), one recovers the strict NLP 
expressions in the hard-collinear region.
The $q^2$-dependent inverse moment of the $B$ LCDA is defined as
\begin{equation}
\label{eq:q2moments}
\frac{1}{\lambdaBp} \equiv  \int_{0}^{\infty}d\omega \,
\frac{\phi^B_{+}(\omega)}{\omega-\nminusq -i 0^+} \ .
\end{equation}
Power corrections to $F_L$ from the photon emission off the 
spectator quark cannot be factorized and are parametrized by the 
``symmetry-preserving'', power-suppressed form factor 
$\xi(q^2,v\cdot q)$. For $q^2 \rightarrow 0$ we recover the result 
for the on-shell $B^- \rightarrow \ell \bar{\nu}_\ell \gamma$ form
 factors~\cite{Beneke:2011nf}.

One important difference between the virtual and on-shell photon 
case concerns the second term in the square brackets in~\eqref{eq:propexpansion}, which matches onto a hadronic matrix element with a 
transverse derivative acting on the spectator-quark field.
This term contributes only to the longitudinal form factor $\Fn$ 
and is hence irrelevant in $B \to \gamma \ell \nu$. Since we 
do not consider explicitly the tree-level contributions 
proportional to the three-particle LCDAs of the $B$ meson, we 
can compute this term in the so-called Wandzura-Wilczek (WW) 
approximation \cite{Beneke:2000wa}, in which case only the 
subleading two-particle LCDA $\phi^B_-(\omega)$ appears. We then
find, for hard-collinear $q^2$,  
\begin{align}\label{eq:ffpara}
F_{A_\parallel}^{\rm NLP} & =  -\frac{2F_B Q_u}{\nplusq}
\left(2m_B\frac{\nminusq}{\nplusq}\frac{1}{\lambdaBm } +1\right) 
+ \frac{2F_B  (Q_b-Q_\ell)}{\nplusq} + \xi'(q^2, v\cdot q)  
\nonumber \\
& =-\frac{4 F_Bm_B Q_u}{(\nplusq)^2} \frac{\nminusq}{\lambdaBm } 
+ \xi'(q^2, v\cdot q)\,.
\end{align}
In the numerical analysis, we employ an expression for 
$F_{A_{\parallel}}$, which is accurate in both the hard and 
hard-collinear $q^2$ region. To this end, we use 
\eqref{eq:redefff} together with 
\begin{align}
\tilde{F}_{A_{\parallel}}^{\rm NLP} & =
\frac{4 F_B m_B Q_u}{\nplusq}\frac{\nminusq}{\nplusq}\left(\frac{1}{\lambdaBp}-\frac{1}{\lambdaBm}\right) -\frac{2 F_B Q_u}{\nplusq}\left(1 + \frac{\nminusq}{\lambdaBp}
\right) \nonumber \\
&+\frac{2 F_B m_b Q_b}{2v\cdot q m_b - q^2} -\frac{2f_B Q_{\lep}}{2v\cdot q}+\xi'(q^2, v\cdot q)\,, 
\end{align}
and  $\Fa^{\rm NLP}$ computed from (\ref{eq:interpolL}), 
(\ref{eq:interpolR}). 
The inverse moment $\lambdaBm$ of the subleading-twist LCDA 
$\phi^B_-(\omega)$, which was already introduced for $B\to K^*\ell\ell$ \cite{Beneke:2001at}, is defined in analogy to~\eqref{eq:q2moments}. 
The finite invariant mass of the virtual photon regulates its 
endpoint divergence at $\omega \to 0$. Nevertheless, in the limit 
$q^2 \to 0$ we find $F_{A_\parallel} \to 0$ due to the additional 
$\nminusq$ in the numerator, as it should be, since an on-shell 
photon has no longitudinal polarization. As in the case of $F_L$ we 
allow for a possible non-factorizable contribution by adding 
an unknown form factor $\xi'(q^2, v\cdot q)$, which must 
also vanish as $q^2\to 0$.

The power-suppressed form factor $\xi$ that parameterizes 
the contribution from soft distances $x\sim 1/\Lambda_{\rm QCD}$ 
between the currents in $T^{\mu\nu}$ as well as 
the three-particle $B$ LCDA contributions have been calculated 
with light-cone QCD sum rules \cite{Wang:2016qii, Beneke:2018wjp}, 
but this method can only be used for $q^2=0$ or space-like. We 
therefore follow the simple ansatz \cite{Beneke:2020fot}
\begin{align}\label{eq:xidef}
\xi(q^2,v\cdot q) & = - r_{\rm LP} \frac{F_B m_B Q_u}{\nplusq}
\frac{1}{\lambdaBp} \nonumber \\
\xi^\prime(q^2,v\cdot q) & = 0 \,,
\end{align}
which incorporates the observation that the power-suppressed form 
factors appear to reduce the LP ones by setting them to a fraction 
$r_{\rm LP}=0.2$ of $F_L$ at tree level. Since there is no LP 
contribution to  $\Fn$, $\xi^\prime$ is set to 0 
in this model. The branching fraction of the four-lepton decay is 
quite sensitive to the value of $r_{\rm LP}$. For the
$B_s\to\mu^+\mu^-\gamma$ decay the conservative estimate 
$r_{\rm LP}=0.2\pm 0.2$ was adopted \cite{Beneke:2020fot}. 
Below we also present results for $r_{\rm LP}=0.2\pm 0.1$.

Since the $B\to\gamma^*$ form factors are time-like, the 
heavy-quark / large-energy expansion is certainly upset locally 
by the lowest light-meson resonances, $\rho$ and $\omega$. 
However, as shown in  \cite{Beneke:2020fot}, quark-hadron 
duality is also violated globally, such that for any $q^2$ bin 
that contains these resonances, the resonance contribution 
will be dominant. 
In order to describe the form factors in the entire region 
$q^2\lesssim 6\,\mbox{GeV}^2$, we add the resonant process  
$B^- \to  \ell \bar{\nu}_\ell V \to \ell \bar{\nu}_{\ell} 
\ell^{(\prime)} \bar{\ell}^{(\prime)}$, where $V=\rho,\omega$,   
to the factorization expressions \eqref{eq:interpolL} and
\eqref{eq:interpolR}. 
As discussed in \cite{Beneke:2020fot}, this procedure can be 
justified parametrically, as the averaged resonance contribution 
is {\em formally} a power correction. Nevertheless, the existence 
of resonances implies that the QCD factorization calculation
of the time-like form factors is not on as solid ground as at 
$q^2=0$  for $B^-\to\ell^-\bar{\nu}_\ell\gamma$. Writing the 
dispersion relation in $q^2$ at fixed $n_+ q$ for the hadronic 
tensor $T^{\mu\nu}$, and including only the $\rho$ and 
$\omega$ resonances in the spectral function in the Breit-Wigner 
approximation, we find 
\begin{align} 
\label{eq:rescon}
F_{L (R)}^{\text{res}} & = \sum_{V=\rho^0\,,\,\omega}{\rm BW}_V 
\,\frac{1}{2}\left(\frac{2m_B}{m_B+m_V}V^{B\to V}(k^2) \pm \frac{m_B+m_V}{v\cdot q} A_1^{B\to V}(k^2)\right) \ , 
\end{align}
where the upper (lower) sign applies to $F_L \,(F_R)$.
In addition, 
\begin{equation}
\label{eq:BW}
{\rm BW}_V\equiv c_{\rm V} \frac{f_{V}m_V}{m_V^2-q^2-i m_V \Gamma_V} \,,
\end{equation}
with $c_{\rho}=1/2$ and $c_\omega=1/6$.\footnote{Compared to 
\cite{Beneke:2020fot}, $c_\rho$ has opposite sign because it 
arises from the $u\bar{u}$ content of the $\rho$ meson, while in 
\cite{Beneke:2020fot} the $d\bar{d}$ component was the relevant 
one.} For the $B\rightarrow V$ transition form factors $V, A_1$ 
and $A_2$ we use the definition and numerical results 
of \cite{Straub:2015ica}. It follows from 
the heavy-quark symmetry relations for the heavy-to-light 
$B\to V$ form factors \cite{Beneke:2000wa}, 
(applicable since $k^2 = k^2(n_+q, q^2)$ via (\ref{eq:qplusk2}) 
and $n_+q \gg \Lambda_{\rm QCD}$ and $q^2=m_V^2$ in the argument 
of the form factors in (\ref{eq:rescon})) 
that  $F_R^{\rm res}$ is power-suppressed relative to $F_L^{\rm res}$, hence it formally counts as a 
next-to-next-to-leading power correction. 
We do not add a resonance contribution to 
the form factor  $F_{A_\parallel}$ for the 
longitudinal intermediate polarization 
state, since a simple Breit-Wigner 
ansatz as above would lead to a non-vanishing form factor 
at $q^2=0$ resulting in a $1/q^4$ singularity in the rate, 
which is unphysical.\footnote{We note that such an ansatz has 
been used in \cite{Danilina:2018uzr, Danilina:2019dji}. We remark 
that $F_{L (R)}^{\text{res}}$ retains an unphysical imaginary 
part at $q^2=0$ from \eqref{eq:BW}, which, however, is even further 
suppressed by the small width $\Gamma_V/m_V\ll 1$ of the resonances.}


\section{Numerical results}

We combine the form factors at leading-power (LP) and 
next-to-leading power (NLP) calculated with QCD factorization 
with the resonance contribution into the final result 
\begin{equation}\label{eq:fftot}
	F_X =  F_X^{\rm LP} +F_X^{\rm NLP}  + F_X^{\rm res} \ .  
\end{equation}
We include renormalization group evolution to sum logarithms 
of the ratio of the hard, hard-collinear and soft scales
in the LP term following \cite{Beneke:2011nf}, but not in 
$F_X^{\rm NLP}$ where we set $F_B=f_B$. Contrary to \cite{Beneke:2020fot}, we do not re-expand 
products of series expansions in $\alpha_s$.

\begin{table}[t]
	\begin{center}
		\begin{tabular}{ c|c||c |c } 
			parameter & value & parameter & value \\    
			\hline \hline
			$|V_{ub}|$ &  $3.70\cdot 10^{-3}$ \cite{Amhis:2019ckw}& $\alem$ & $1/132.18$ \\
			$m_b$ & $4.78$ GeV & $f_B$ & $190 \,\,\text{MeV}$ \cite{FLAG}\\
			$m_{\rho}$ & $770 \,\, \text{MeV} $ & $m_{\omega}$ & $782 \,\, \text{MeV}$  \\ 
			$\Gamma_{\rho}$    &  $147.8\,\,  \text{MeV}$ & $\Gamma_{\omega}$ & $8.49 \,\, \text{MeV}$   \\ 
			$f_{\rho}$ & $213 \,\, \text{MeV}$ \cite{Straub:2015ica}  & $f_{\omega}$ &$197  \,\, \text{MeV}$ \cite{Straub:2015ica}
		\end{tabular} 
	\end{center}
	\caption{Input parameters from PDG \cite{pdg} unless stated otherwise. The value of $f_B$ is taken from FLAG \cite{FLAG} using inputs from \cite{Bussone:2016iua, Bazavov:2017lyh, Hughes:2017spc, Dowdall:2013tga}. Here we quote the exclusive $|V_{ub}|$ value from HFLAV \cite{Amhis:2019ckw}, which uses lattice inputs from \cite{ Lattice:2015tia, Flynn:2015mha}. Here $\alem=\alem^{(5)}(5 \;\rm{GeV})$. }
	\label{tab:inputs}
\end{table}

We use the inputs specified in Table \ref{tab:inputs} and the 
exponential model
\begin{equation}
\label{eq:expmod}
	\phi^B_{+}(\omega) =\frac{\omega}{\omega_0^2}e^{-\omega/\omega_0} \ , \quad \quad \phi^B_{-}(\omega) =\frac{1}{\omega_0}e^{-\omega/\omega_0} \ ,
\end{equation}
for the $B$ LCDA. We put $\omega_0 =  \lambda_B \equiv \lambda_B^+(n_-q=0) = 0.35\pm 0.15$ GeV at the scale 1~GeV as our default value. In the LP terms, we evolve  $\phi^B_+(\omega)$ to the hard-collinear scale $\mu_{hc}$ employing the analytic expression 
given in \cite{Beneke:2018wjp}. 
Previous analyses of $B\to \gamma \ell \nu$ showed that the shape of the $B$ LCDA is also important when including power corrections \cite{Beneke:2018wjp}. For the time-like virtual photon 
form factors, there is less control over power corrections 
and we therefore content ourselves with the exponential model to present our main results and conclusions. We further study the dependence of $\lambda_B^\pm(n_-q)$ and the branching fraction of the four-lepton decay on the shape of the $B$ meson LCDA in Sec.~\ref{sec:shapeanalysis} using three two-parameter models \cite{Beneke:2018wjp} for the $B$ LCDA. 
In addition, as $|V_{ub}|$ is an overall factor we do not include its uncertainty in our error estimates, nor do we include the negligible uncertainties on the other input parameters in Table~\ref{tab:inputs}.  
We expect that eventual measurements of the four-lepton final 
states will be normalized to the decay rate of another, 
accurately known, exclusive $b\to u$ transition. 

\subsection{Form factors}

\begin{figure}[t]
	\centering
	\subfloat{\includegraphics[scale=0.85]{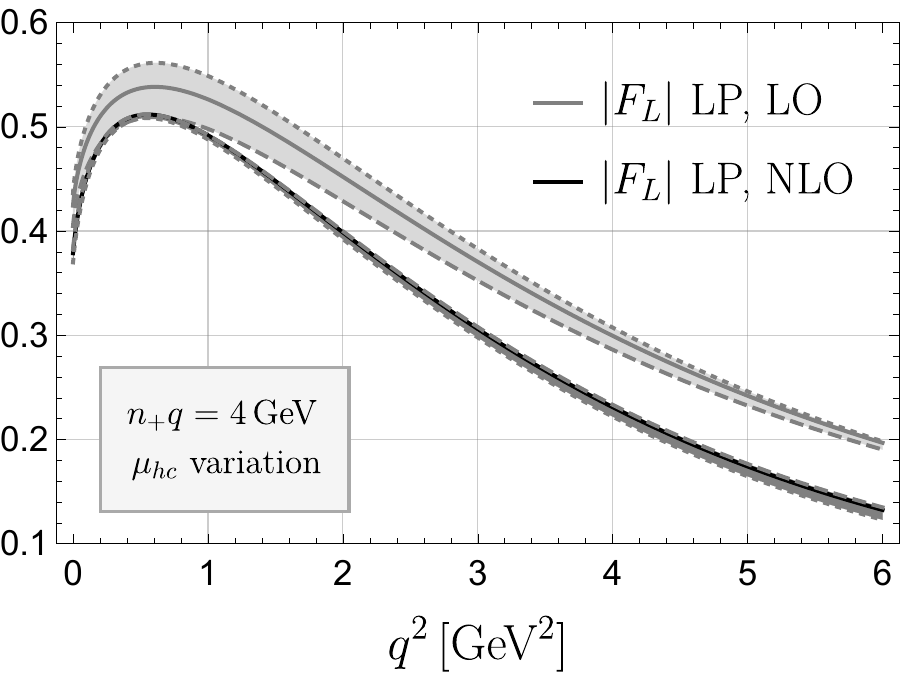}}\hfill
	\subfloat{\includegraphics[scale=0.85]{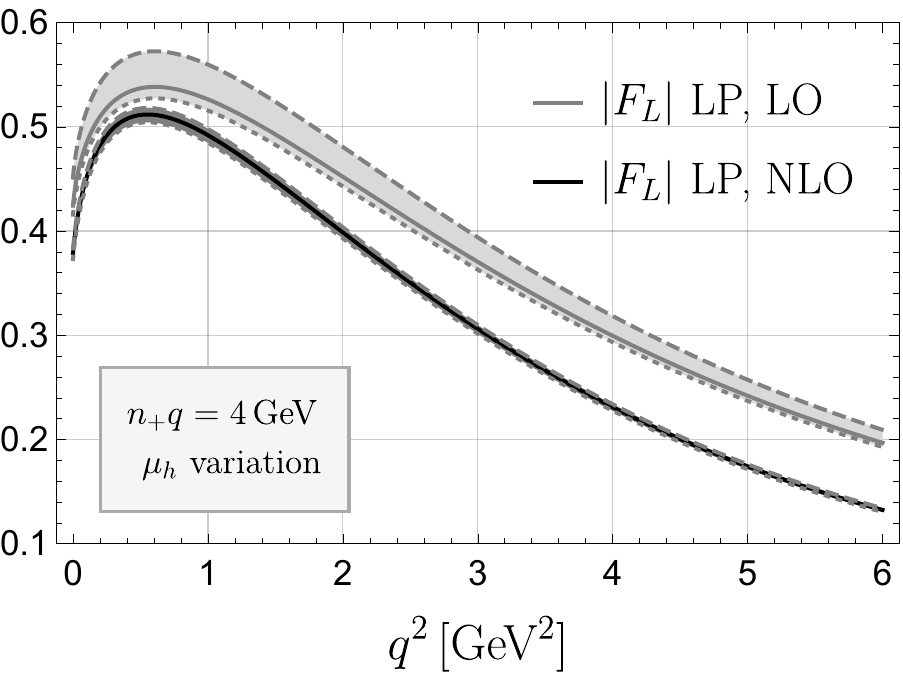}}
\caption{$|F_L^{\rm LP}|$ at leading order (LP, LO) and 
		next-to-leading order (LP, NLO) as a function of $q^2$. The bands represent the scale uncertainty from $\mu_{hc}=1.5\pm0.5$~GeV (left) and $\mu_h=5^{+5}_{-2.5}$~GeV (right) and the dashed (dotted) curves correspond to the upper (lower) scale value. The value 
		of $n_+q$ is fixed to 4~GeV.}
	\label{fig:scaledep}
\end{figure}

In Fig.~\ref{fig:scaledep}, 
we show $|F_L^{\rm LP}|$ at 
leading order (LP,LO) and next-to-leading order (LP,NLO) 
as a function of $q^2$ at fixed $\nplusq=4$ GeV. The band describes the scale uncertainty of the hard-collinear scale $\mu_{hc}=1.5\pm0.5$ GeV (left) and that of the hard scale $\mu_h = 5^{+5}_{-2.5}$ GeV (right). Similar to the $B\to \gamma \ell \nu$ case, at small 
$q^2$ the form factor in the LO approximation has a large scale uncertainty, which 
is practically removed  by the NLO correction. We conclude that the 
LP form factor is under very good control away from 
light-meson resonances, once the $B$ LCDA 
input is specified. It is worth noting that the 
form factors do not fall off right away with increasing 
$q^2$, but exhibit a maximum near $q^2 \approx 0.5\,$GeV$^2$. 
The maximum is generated by the sizeable imaginary part 
$\pi\phi^B_+(n_-q)$ of the $q^2$-dependent $B$ LCDA 
moment $\lambda_B^+(n_-q)$. These features of $F_L$ are 
largely independent of the chosen value of $n_+ q$.

\begin{figure}[t]
	\centering
	\includegraphics[scale=0.85]{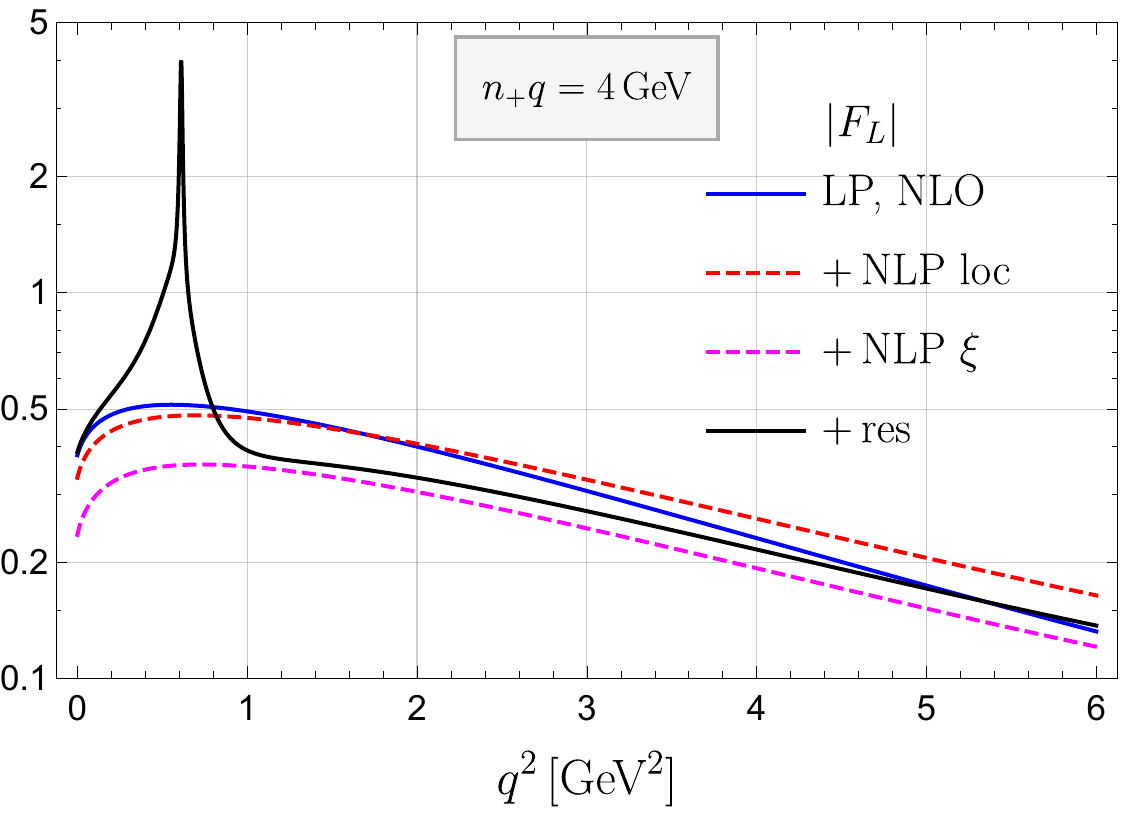}
	\caption{Illustration of the $q^2$ dependence of the leading form factor $|F_L|$ including successively leading power (LP, NLO), next-to-leading power (NLP) local contributions, $\xi$ and resonances.  The value of $n_+q$ is fixed to 4~GeV.}
	\label{fig:FLsplit}
\end{figure}

The breakdown of $|F_L|$ into its various contributions is 
shown in Fig.~\ref{fig:FLsplit}, 
starting with LP,NLO, then successively adding the NLP local (loc) contributions (defined as  $F_X^{\rm NLP}$ without the $\xi$ term), the $\xi$-contribution as defined in \eqref{eq:xidef} and finally the resonance contribution (res). We observe that the NLP contribution is of similar size as the NLO correction at LP. In the small $q^2$ region, the form factor is locally dominated by the resonance contribution, as expected. However, also at larger $q^2$ the resonance contribution is comparable to the NLP local contribution. This is due to the fact that the fall-off of the form factors in QCD factorization with increasing 
$q^2$ is not faster than the $1/q^2$ fall-off 
of the Breit-Wigner parametrization of the resonances. Note that we have fixed again $\nplusq=4$~GeV, and only show the $q^2$ dependence of the form factors as the above observations are generic.

\begin{figure}[t]
\centering
\includegraphics[scale=0.9]{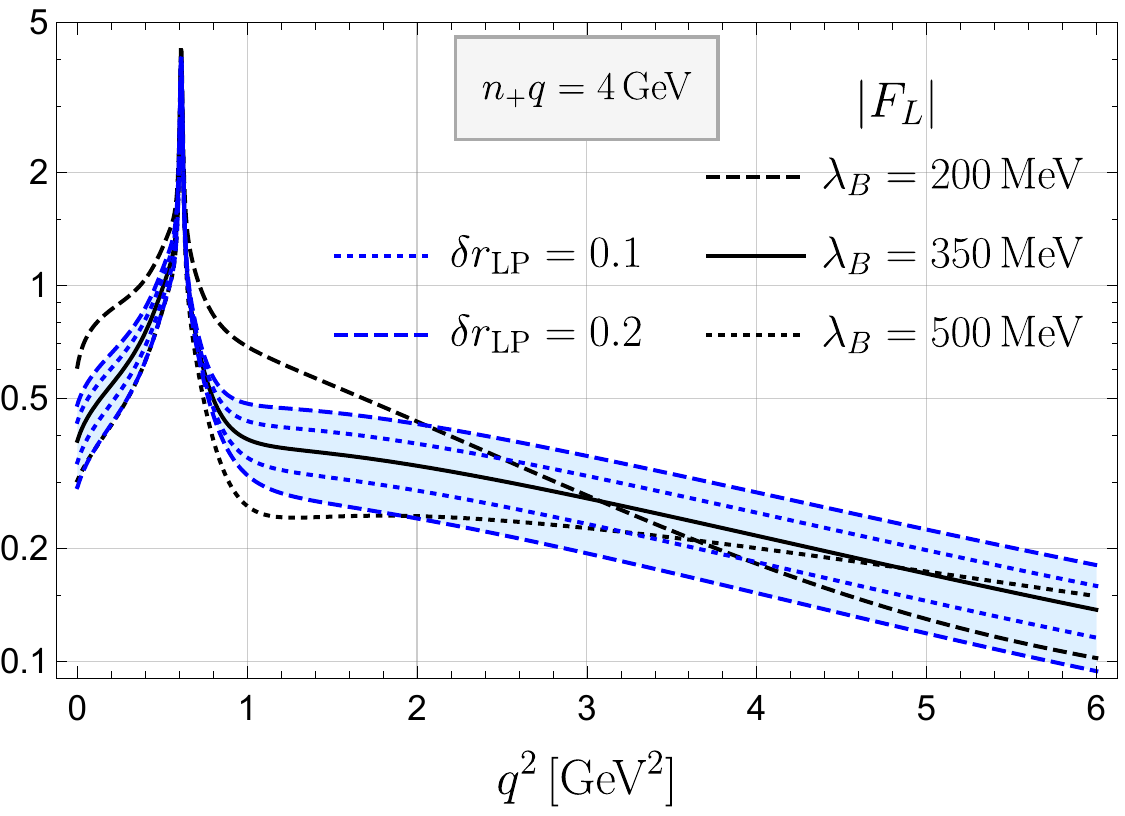}\\[0.4cm]
\includegraphics[scale=0.75]{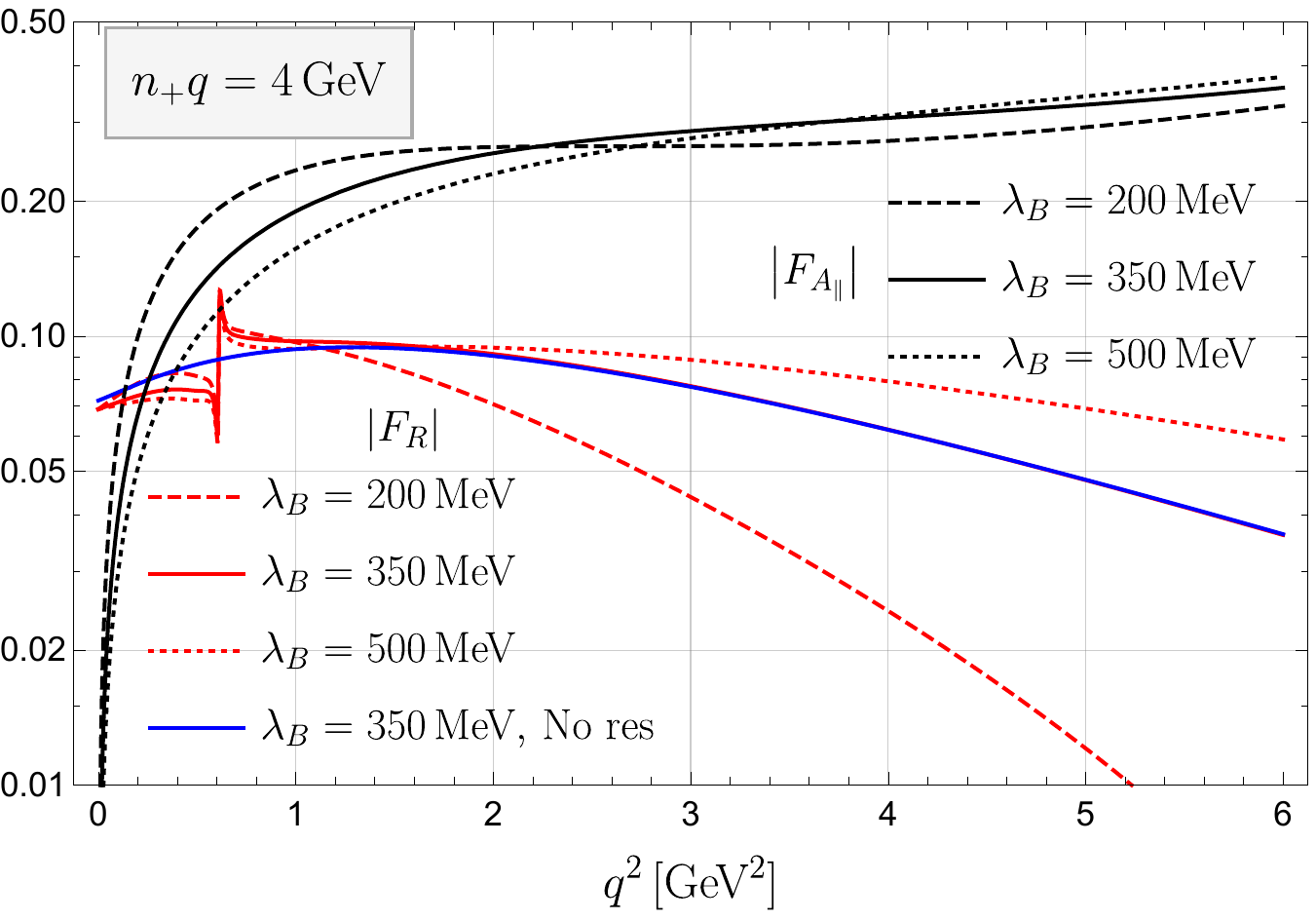}
\caption{The $q^2$ dependence of the form factors at fixed $\nplusq = 4$ GeV for three values of $\lambda_B$ from $200$ MeV (dashed) to $500$ MeV (dotted). In addition, for $|F_L|$ (upper panel) we show the uncertainty from varying for the central value $r_{\rm LP}$ by $\delta r_{\rm LP} = 0.1 (0.2)$. For the NLP form factors (lower panel), we show $|F_R|$ for both, with and without the resonance contribution. We do not add a resonance 
contribution to $F_{A_\parallel}$.}
	\label{fig:errorbudget}
\end{figure}

The $q^2$ dependence of the power-suppressed (NLP) form factors $F_R$ and $F_{A_\parallel}$ is shown in the lower panel of Fig.~\ref{fig:errorbudget}. For $F_R$, we show separately the local NLP contribution and the total by adding the resonance contribution. As we do not include a resonance contribution for $F_{A_\parallel}$, we only show the total form factor. 
In addition, we show the dependence on $\lambda_B$ by varying it from $200$ MeV (dashed) to $500$ MeV (dotted). 
Except for very small $q^2$, the form factor relevant to the 
longitudinal polarization state of the virtual photon is 
significantly larger than the one for the right-helicity state.

For $F_L$ (upper panel of Fig.~\ref{fig:errorbudget}), we show in addition to the results for  $\lambda_B=200$ ~MeV (dashed) and $500$~MeV (dotted) the dominant uncertainty of $|F_L|$ computed 
with the central value $\lambda_B=350$~MeV from varying $r_{\rm LP}$ by $\delta r_{\rm LP}=0.1 (0.2)$. We note two important features. First, for all three form factors there is a crossing of the dashed and dotted lines, such that the lower value $\lambda_B = 200$ MeV increases the form factors at small $q^2$ but decreases it for 
large $q^2$, while for the upper value $\lambda_B = 500$ MeV the situation is reversed. In the region where the crossing occurs (around $3.5$~GeV$^2$ for $F_L$) all sensitivity to $\lambda_B$ is lost. Second, for $F_L$ at low $q^2$ the sensitivity to $\lambda_B$ is larger than the uncertainty coming from $r_{\rm LP}$. 

\begin{figure}
	\centering
	\includegraphics[scale=0.85]{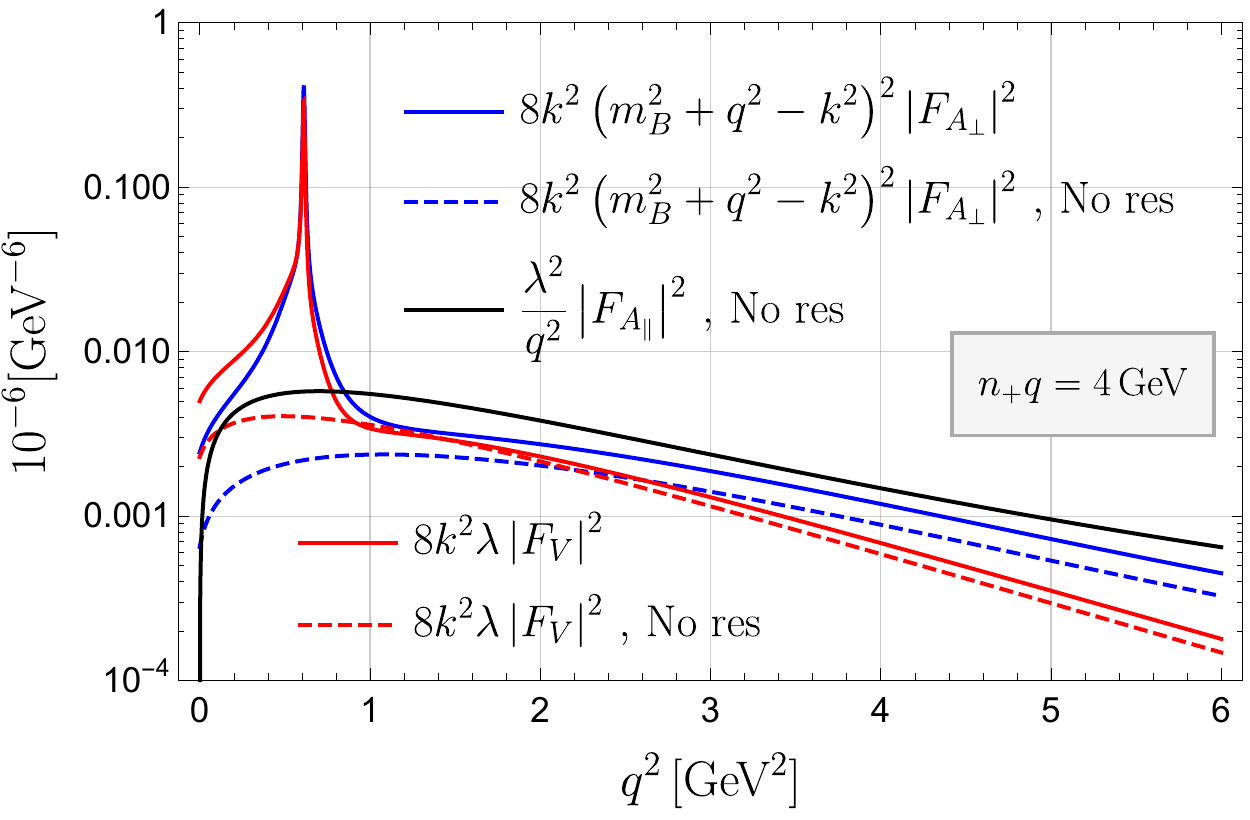}
	\caption{Contribution of the form factors to the rate in \eqref{eq:d2Br} including kinematic factors. For $F_V$ and $F_{A_\perp}$, we show the QCDF results (labelled ``No res'') and the full result including resonances.  }
	\label{fig:ratecontri}
\end{figure}

Finally, we comment on the contribution of the three form factors to the differential rate in \eqref{eq:d2Br}. More precisely, we show in 
Fig.~\ref{fig:ratecontri} the three terms in the 
round bracket in \eqref{eq:d2Br}, that is, the form factors 
squared including their kinematic prefactors. It is 
remarkable that the longitudinal polarization term 
$\frac{\lambda^2}{q^2}\,|F_{A_\parallel}|^2$ dominates 
the rate outside the resonance region, despite that 
fact that it is technically power-suppressed. Moreover, 
leaving out the resonance term, the longitudinal term 
would dominate even at small $q^2$, although it vanishes 
for $q^2\to 0$ (since $F_{A_\parallel} \sim q^2$ as 
$q^2\to 0$), while the other two terms approach constants 
in this limit.

This behaviour can be understood by comparing the analytic 
expressions for the three terms (without the resonance term) 
for small $q^2$. For small $q^2$, the first two terms 
in the round bracket of \eqref{eq:d2Br}
combine to $16 k^2 (m_B^2-k^2)^2 (|F_L|^2+|F_R|^2)$, 
and we then estimate 
\begin{eqnarray}
	\frac{d^2 {\rm Br}^{(F_{A_{\parallel}})}}{dq^2\,dk^2}
	\mbox{\Huge$\mathbin{/}$}
	\frac{d^2 {\rm Br}^{(F_L)}}{dq^2\,dk^2} 
	&=& \frac{m_B}{m_B-n_+q}\frac{q^2}{(n_+ q)^2} 
	\left(\ln^2\frac{q^2 e^{\gamma_E}}{n_+q \lambda_B^+}+\pi^2\right) 
	+\mathcal{O}\!\left(\frac{q^4}{m_B^4}\right)
	\nonumber\\
	&\approx& \frac{27 \pi^2 q^2}{4 m_B^2} 
	+\mathcal{O}\!\left(\frac{q^4}{m_B^4}\right)\,,
\end{eqnarray}
where the last line refers to the representative value  
$\nplusq=2 m_B/3$. The parametric dependence identifies 
this ratio as power-suppressed in the hard-collinear 
region $q^2\sim m_B \Lambda_{\rm QCD}$ as it should be. 
However, the large numerical factor $27\pi^2/4$ implies 
that the longitudinal term dominates whenever 
$q^2$ is larger than the very small value $0.4~$GeV$^2$ 
as seen in the Figure. The origin of the large factor 
is the $\pi^2$ that arises from the large imaginary 
part of the inverse $B$ LCDA moment, in this case 
$\lambda_B^-(n_-q)$ in (\ref{eq:ffpara}), since for values 
$q^2 \in [0.1,1]$ the logarithmic term 
$\ln^2\frac{q^2 e^{\gamma_E}}{n_+q \lambda_B^+}$
is small. 

\subsection{Predictions for the branching ratios}

In this section, we provide theoretical predictions for the 
branching ratio in various $q^2$ bins, integrated over 
$n_+q$ (alternatively, $k^2$). The factorization calculation 
of the form factors in \eqref{eq:lpFF}, \eqref{eq:interpolL}, \eqref{eq:interpolR} and \eqref{eq:ffpara} are valid only 
for $\nplusq\sim \mathcal{O}(m_B)$. We therefore assume 
$\nplusq > 3$~GeV, which corresponds to $E_\gamma =1.5$ GeV 
at $q^2=0$, 
and integrate the double-differential branching fraction 
over $\nplusq>3$~GeV before forming $q^2$ bins. A rough estimate, obtained by assuming that our results apply in the full phase space, shows that the $\nplusq$ cut reduces the rate by $\mathcal{O}(20\%)$ for the $[1.5,6]$ GeV$^2$ $q^2$ bin.   

For non-identical lepton flavours, $\ell^\prime \neq \ell$, the required $\nplusq$ cut can easily be applied as for each event $\nplusq$ can be inferred from the reconstructed $k^2$ and $q^2$ using \eqref{eq:qplusk2}. For the $q^2$ bins, we consider the low bin $[4m_\mu^2,0.96]$ GeV$^2$, where the upper boundary of the bin is determined such that the large experimental background from $\phi$ mesons decaying into a  lepton pair is avoided. This bin was also considered by the LHCb Collaboration \cite{Aaij:2018pka}. Fig.~\ref{fig:errorbudget} shows that in this bin, the $\rho$ and $\omega$ resonances make a large contribution. As mentioned, we do not attribute an additional error due to our resonance model. This introduces an additional uncertainty in this region which is challenging to quantify. Above $q^2>1$ GeV$^2$, the effect of the $\rho$ and $\omega$ resonances (and thus a possible uncertainty associated with this) is significantly reduced. We consider three different $q^2$ bins: $[1,6], [1.5,6]$ and $[2,6]$ GeV$^2$. In these bins, the resonance contribution is approximately 10\% only. In Table~\ref{tab:BRdifflep}, we give the branching ratio in these $q^2$ bins, specifying the contributions which are successively added. In addition, we specify the uncertainties from variations of the scales $\mu_{h,hc}$, $r_{\rm LP}=0.2\pm0.2$ and  $\lambda_B=350\pm150$ MeV. We observe that in the three considered regions above $q^2>1$ GeV$^2$, the effect of the resonances is smaller than the uncertainty from~$r_{\rm LP}$. 

\floatsetup[table]{font=small}


\begin{table}[t]
	\begin{center}
		\begin{tabular}{ |c|c|c c|c c | c |c c c c|}
			\hline 
			Decay & $q^2$ bin  & \multicolumn{2}{|c|}{LP} & \multicolumn{2}{|c|}{NLP} & Total  &\multicolumn{4}{|c|}{Uncertainty} \\    
			& $[\text{GeV}^2]$ & LO & NLO & loc & +$\xi$ & +res & $\mu_{h,hc}$ & $r_{\text{LP}}$ & $\lambda_B$ & tot \\ 
			\hline 
			
			\multirow{5}{5em}{$\mu^- \mu^+\,e^- \,\bar{\nu}_{e} $} 
			
			& $[4m_{\mu}^2,0.96]$ 
			&0.58 &0.51 
			&0.70 & 0.48 & 1.57 
			&$^{+0.02}_{-0.02}$ &$^{+0.35}_{-0.29}$ &$^{+1.33}_{-0.40}$ &$^{+1.37}_{-0.49}$ \\
			
			& $[4m_{\mu}^2,6]$ 
			&0.76 &0.66 
			&0.98 & 0.67 & 1.78 
			&$^{+0.02}_{-0.02}$  &$^{+0.43}_{-0.35}$
			&$^{+1.46}_{-0.47}$& $^{+1.52}_{-0.58}$ \\
			
			& $[1,6]$
			&0.18 &0.14 
			&0.26 & 0.18 & 0.20 
			&$^{+0.00}_{-0.00}$  &$^{+0.08}_{-0.06}$ 
			&$^{+0.11}_{-0.06}$
			&$^{+0.14}_{-0.08}$ \\
			
			& $[1.5,6]$
			&0.10 &0.08 
			&0.15 & 0.10 & 0.11 
			&$^{+0.00}_{-0.00}$ &$^{+0.05}_{-0.04}$ &$^{+0.03}_{-0.03}$ 
			&$^{+0.06}_{-0.05}$ \\
			
			& $[2,6]$
			&0.062 &0.042 
			&0.090 & 0.062 & 0.068
			&$^{+0.001}_{-0.001}$  &$^{+0.030}_{-0.022}$ &$^{+0.002}_{-0.012}$
			&$^{+0.030}_{-0.025}$ \\
			
			\hline
			\multirow{2}{5em}{$e^-e^+ \mu^- \bar{\nu}_{\mu}$} 
			& $[q^2_{\rm min},0.96]$ 
			&1.23 &1.04 
			&1.23 & 0.81 & 2.28 
			&$^{+0.03}_{-0.04}$ &$^{+0.66}_{-0.53}$ &$^{+2.40}_{-0.67}$ &$^{+2.49}_{-0.86}$ \\
			
			& $[1,6]$ 
			&0.18 &0.14 
			&0.26 & 0.18 & 0.20 
			&$^{+0.00}_{-0.00}$ &$^{+0.08}_{-0.06}$ &$^{+0.11}_{-0.06}$ &$^{+0.14}_{-0.08}$ \\
			
			\hline
		\end{tabular}
	\end{center}
	\caption{Branching ratio for the two non-identical lepton flavour cases (in $10^{-8}$) integrated over different bins in $q^2$ and for $\nplusq> 3$ GeV. We show the individual contributions consecutively adding to the LP result the NLP local and $\xi$ contributions and finally the resonances. In addition, we quote the uncertainties from varying the scales $\mu_{h,hc}$,  \mbox{$r_{\rm LP}=0.2\pm0.2$} and $\lambda_B=350\pm150$ MeV. The total uncertainty is obtained by adding them in quadrature. For electrons, we also consider a low bin with $q^2_{\rm min}= 0.0025$~GeV$^2$. }
	\label{tab:BRdifflep}
\end{table}


\subsubsection{Identical lepton flavours}

A challenge arises when considering identical lepton flavours, $\ell^\prime = \ell$, because experimentally the two like-sign leptons cannot be distinguished. This results in the additional interference term \eqref{eq:brident}. More challenging is the required cut on $\nplusq$, where $q$ is the photon momentum, to ensure that the photon has hard-collinear momentum. 
Considering again $B^-\to \ell^-(p_1) \ell^+(p_2)\ell^-(p_3) \bar{\nu}_\ell(p_\nu)$, with $q^2=(p_1+p_2)^2$ and $\tilde{q}^2=(p_2+p_3)^2$, 
experimentally, $q^2$ and $\tilde{q}^2$ cannot be distinguished. 
Instead, the invariant mass of two $\mu^- \mu^+$-pairs are defined as $q^2_{\rm low}<q^2_{\rm high}$. In this case, placing the required cut on $\nplusq$ is not unambiguously possible as we cannot determine if the virtual photon has $q^2_{\rm low}$ or $q^2_{\rm high}$ associated with its momentum. To deal with this issue, several observations can be made:
\begin{itemize}
	\item for small $q^2_{\rm low}$, the photon momentum can be associated with $q^2_{\rm low}$ most of the time. If this is the case, a cut on $\nplusq_{\rm low}> 3$ GeV suffices (similar to the non-identical lepton flavour case). In fact, a more detailed analysis shows that the cases falling outside this cut (i.e. the region which cannot be described in QCD factorization in which the photon has $q^2_{\rm high}$ but $\nplusq$ small) is phase-space suppressed by two 
powers of $1/m_b$ compared to the leading contribution. 
	\item for $q^2$ bins above $1$ GeV$^2$, the situation is more complicated as the photon more often has $q^2_{\rm high}$. Therefore, we have to ensure $\nplusq>3~$GeV for both $q^2_{\rm low}$ and $q^2_{\rm high}$. 
\end{itemize}

We thus have to restrict both $n_+q_{\rm low}> 3$ GeV and $n_+q_{\rm high}> 3$ GeV. These quantities are now defined in the following way:\footnote{
We remark that, unlike previously, here $n_+ q_{\rm high}$ does not coincide with the component of $q^\mu_{\rm high}$ in the $n_-^\mu$ direction as defined above~\eqref{eq:qplusk2}, if the momentum of the $\gamma^*$ is $q_{\rm low}^\mu$. The reason is that we always align the $z$-axis with the three-momentum of the $\gamma^*$, but we do not know which of $q_{\rm low}^\mu$ and $q_{\rm high}^\mu$ refers to the 
virtual photon momentum.} 
For each event, we specify $q^2_{\rm low}$ and $q^2_{\rm high}$. We can then associate the remaining lepton plus neutrino as $k^2_{\rm low}$ and $k^2_{\rm high}$, respectively. Here high and low are just labels and in this case  $k^2_{\rm low}$ is not necessarily lower than  $k^2_{\rm high}$. Then using \eqref{eq:qplusk2}, both $\nplusq_{\rm low}$ and $\nplusq_{\rm high}$ can be calculated from their corresponding $k^2$ and $q^2$. Alternatively, one could cut on $k^2_{\rm low}$ and $k^2_{\rm high}$ directly. 


\begin{table}[t]
	\begin{center}
		\begin{tabular}{ |c|c|c c|c c|c|c c c c|} 
			\hline
			Decay & $q_{\rm low}^2$ bin  & \multicolumn{2}{|c|}{LP} & \multicolumn{2}{|c|}{NLP} & Total & \multicolumn{4}{|c|}{Uncertainty} \\    
			& $[\text{GeV}^2]$ & LO & NLO & loc & +$\xi$ & +res & $\mu_{h,hc}$ & $r_{\text{LP}}$ & $\lambda_B$  & tot \\ 
			\hline 
			\multirow{5}{5em}{$\mu^-\mu^+\mu^- \bar{\nu}_{\mu}$} 
			& $[4m_{\mu}^2,0.96]$ 
			&0.58 &0.51 
			&0.71 & 0.49 & 1.54 {\scriptsize{(1.77)}}
			&$^{+0.02}_{-0.02}$  &$^{+0.35}_{-0.29}$
			&$^{+1.29}_{-0.39}$
			&$^{+1.34}_{-0.48}$ \\ 
			
			& $[4m_{\mu}^2,6]$ 
			&0.74 &0.64 
			&0.97 & 0.67 & 1.75 {\scriptsize{(2.00)}}
			&$^{+0.02}_{-0.02}$ &$^{+0.42}_{-0.34}$ &$^{+1.40}_{-0.45}$ &$^{+1.46}_{-0.56}$ \\
			
			& $[1,6]$
			&0.15 &0.11 
			&0.25 & 0.17 & 0.19 {\scriptsize{(0.21)}}
			&$^{+0.00}_{-0.00}$ &$^{+0.07}_{-0.05}$ &$^{+0.10}_{-0.05}$ 
			&$^{+0.12}_{-0.06}$ \\ 
			
			& $[1.5,6]$
			&0.08 &0.06 
			&0.14 & 0.10 & 0.11 {\scriptsize{(0.11)}}
			&$^{+0.01}_{-0.01}$ &$^{+0.04}_{-0.03}$ &$^{+0.03}_{-0.02}$ &$^{+0.05}_{-0.04}$ \\ 
			
			& $[2,6]$
			&0.04 &0.03 
			&0.08 & 0.06 & 0.06 {\scriptsize{(0.07)}}
			&$^{+0.00}_{-0.00}$ &$^{-0.02}_{-0.02}$ &$^{+0.00}_{-0.01}$ &$^{+0.03}_{-0.02}$ \\ 
			
			\hline
			\multirow{2}{5em}{$e^-e^+e^- \bar{\nu}_{e}$} 
			
			& $[q^2_{\rm min},0.96]$ 
			&1.22 &1.03 
			&1.23 & 0.80 & 2.23 {\scriptsize{(2.57)}}
			&$^{+0.04}_{-0.06}$ &$^{+0.65}_{-0.53}$ &$^{+2.33}_{-0.65}$ &$^{+2.42}_{-0.82}$ \\ 
			
			& $[1,6]$ 
			&0.15 & 0.12
			&0.25  &0.18 & 0.20   {\scriptsize{(0.22)}}
			&$^{+0.00}_{-0.00}$ &$^{+0.07}_{-0.05}$ &$^{+0.10}_{-0.05}$ &$^{+0.12}_{-0.07}$ \\ 
			\hline
		\end{tabular}
	\end{center}
	\caption{Branching ratio for the two identical lepton cases (in $10^{-8}$) integrated over different bins in $q_{\rm low}^2$ applying two cuts: $\nplusq_{\rm low}>3$ GeV and $\nplusq_{\rm high}>3$ GeV. We show the individual contributions consecutively adding to the LP result the NLP local and $\xi$ contributions and finally the resonances. In addition, we quote the uncertainties from varying the scales $\mu_{\rm h,hc}$,  $r_{\rm LP}=0.2\pm0.2$ and $\lambda_B=350\pm150$ MeV. The total uncertainty is obtained by adding these contribution in quadrature. For the total results, we also quote the result with only one cut: $\nplusq_{\rm low}> 3$ GeV in parenthesis. For electrons, we also consider a low bin with $q^2_{\rm min}= 0.0025$~GeV$^2$.}
	\label{tab:BRidlep}
\end{table}

For our final results in Table~\ref{tab:BRidlep}, we thus include two cuts: $n_+q_{\rm low}$ and $n_+q_{\rm high}>3$ GeV for all bins. Our final results for the branching ratio for different $q_{\rm low}^2$ bins are given in Table~\ref{tab:BRidlep}. Again we present the different contributions added successively. We emphasize that placing these two cuts on $\nplusq$ might be conservative, specifically for the low $q^2$ bin as discussed above, given the phase space suppression of the region in which $q^2_{\rm high}$ is associated with the photon. We confirm numerically that indeed this region 
is small, by calculating the rate with and without the cut on $\nplusq_{\rm high}$. For comparison, in Table~\ref{tab:BRidlep} we also give the results for the total rate with only the $\nplusq_{\rm low}$ cut in parenthesis. For identical lepton flavours, the branching ratio contains two contributions as defined in \eqref{eq:brident}. With this convention, we find that Br$_{\rm int}$ contributes positively to the rate but is suppressed by at least one order of magnitude compared to the non-identical lepton flavour rate. 

A comment on the low-$q^2$ bin for $B^-\to \mu^-\mu^+\mu^-\bar\nu_\mu$ is in order. Our prediction for Br($B^+ \to \mu^-\mu^+\mu^-\bar\nu_\mu$) is $1.54 \;(1.77)\cdot 10^{-8}$ and includes cuts on $\nplusq$. 
Yet, it lies close to the upper limit   $<1.6\cdot 10^{-8}$ given by the LHCb collaboration for this decay mode in this bin \cite{Aaij:2018pka}. Hence the LHCb result may already point towards a larger value 
of $\lambda_B$.

\subsection{Sensitivity to $\lambda_B$}

\begin{figure}[t]
\centering
\includegraphics[scale=0.6]{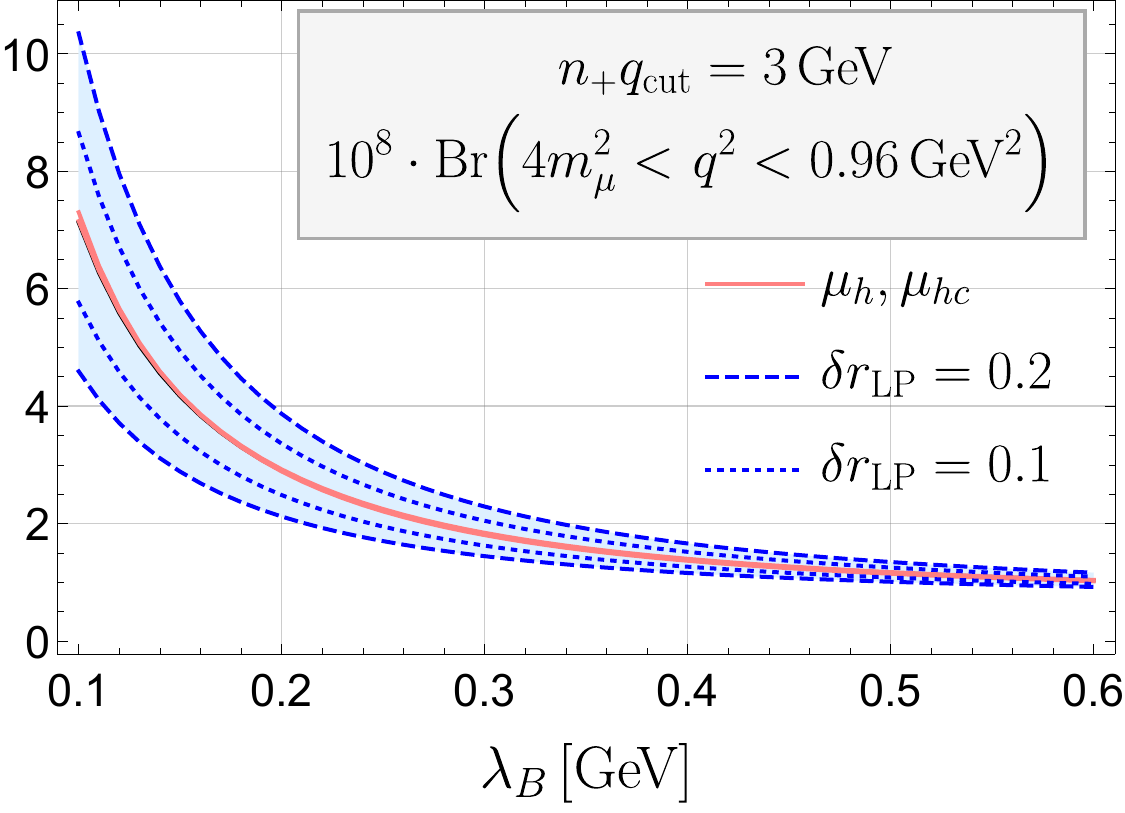}\\[0.2cm]
\includegraphics[scale=0.55]{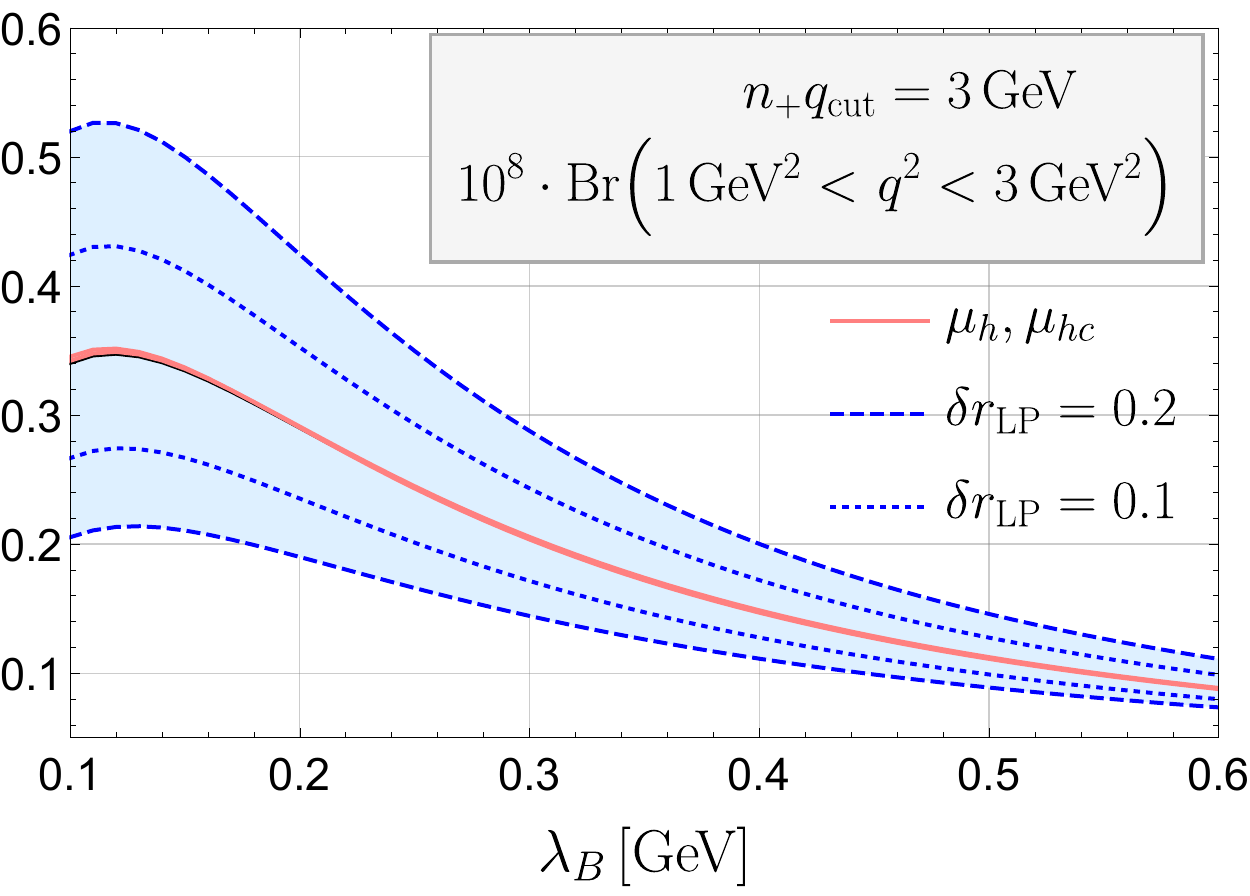}\hskip0.6cm 
\includegraphics[scale=0.55]{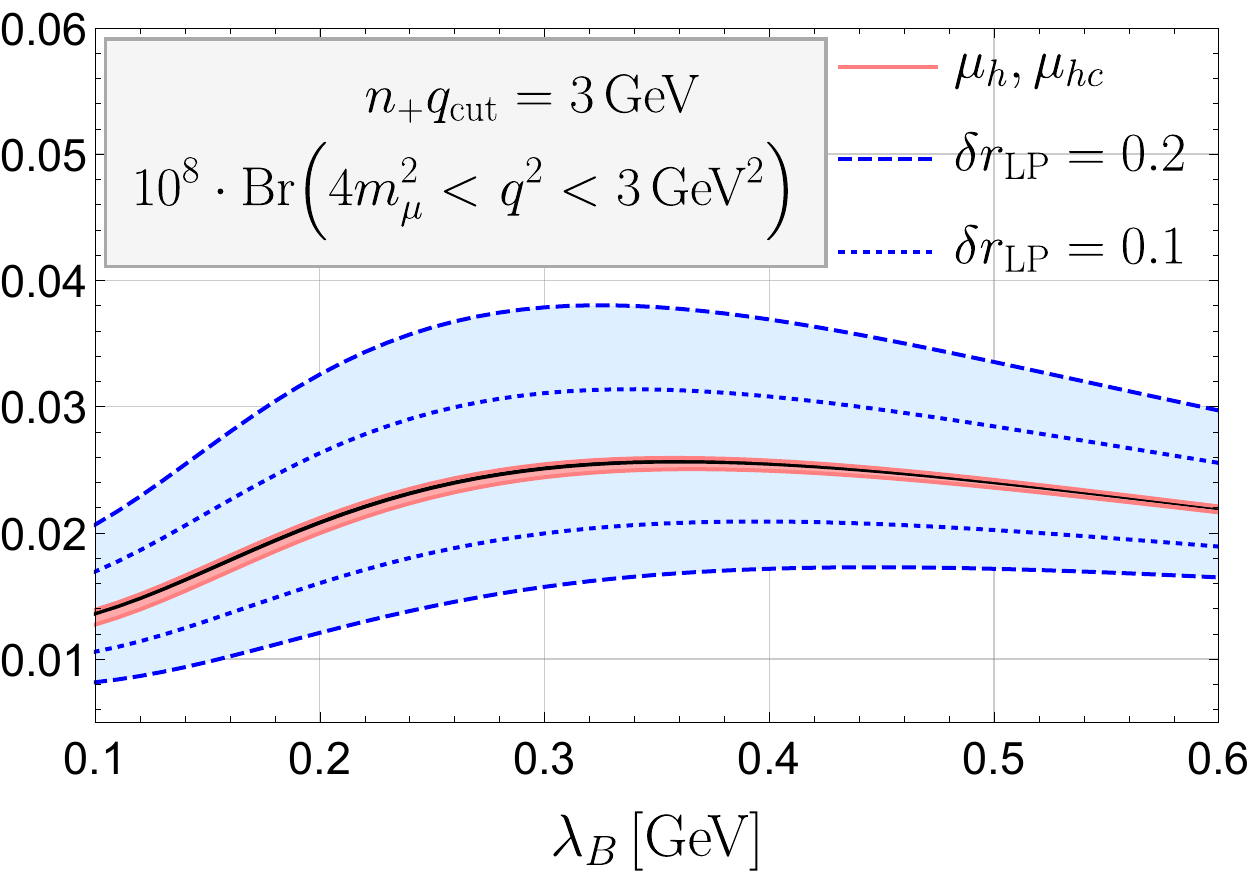}
\caption{ Branching ratio (in $10^{-8}$) for three different $q^2$ bins as a function of $\lambda_B$ for the decay mode $B^-\to \mu^- \mu^+\,e^- \,\bar{\nu}_{e} $. We include the variation from $\mu_{h, hc}$ and from $\delta r_{\rm LP}$ as inner (red) and outer (blue) bands, respectively. }
\label{fig:brlambda}
\end{figure}

Our predictions for the branching ratio suffer from a large uncertainty due to $\lambda_B$. Therefore, a measurement of the branching ratio may be used to obtain a bound on $\lambda_B$. Figure~\ref{fig:brlambda} shows the rate as a function of $\lambda_B$ for the $[4m_\mu^2, 0.96]$ GeV$^2$ bin and for the $[1,3]$ and $[3,6]$ GeV$^2$ bins. We have split the $[1,6]$ GeV$^2$ to avoid integrating over the region where the sensitivity to $\lambda_B$ variation switches sign (see Fig.~\ref{fig:errorbudget}). As the uncertainty of the branching ratio is dominated by the error on $r_{\rm LP}$, we consider two options; $\delta r_{\rm LP}=0.2$ and $\delta r_{\rm LP}=0.1$. The latter option shrinks the error by half. These predictions include our model for the long-distance resonance contributions as described above, for which we do not add an uncertainty. We note that the sensitivity to $\lambda_B$ is best for the small $q^2$ bins, while it is significantly reduced for higher $q^2$ bins. Comparing to $B\to \gamma\ell \nu_\ell$ \cite{Beneke:2018wjp}, we conclude that the sensitivity to $\lambda_B$ for $B\to \ell \nu_\ell \ell^\prime \bar{\ell}^\prime$ in the low-$q^2$ bin is comparable (compare to Fig.~5 in \cite{Beneke:2011nf}) for $\lambda_B<200\,$MeV, but less when it is larger. However, in this bin the resonance contribution is sizeable and there is unquantified model dependence related to its interference with the factorization contribution. For the $[1,3]$ GeV$^2$ bin, the resonance contribution is less pronounced and thus this bin could still provide information on $\lambda_B$ despite its smaller sensitivity. 

\subsection{Dependence on the shape of the $B$ LCDA}
\label{sec:shapeanalysis}

Up to now, we used the exponential model \eqref{eq:expmod} 
to present our main results. However, it is known that for 
$B\to \gamma \ell\nu$ \cite{Beneke:2018wjp} the shape of 
the $B$ meson LCDA has a significant effect through the 
dependence of radiative corrections on the logarithmic 
inverse moments, and of the power-suppressed form factor 
$\xi$ through its dependence on the shape of the LCDA in the 
sum rule calculation. In four-lepton decay the generalized 
inverse moments $\lambda_B^\pm(\nm q)$ introduce further 
dependence on the shape of the LCDA. 

\begin{figure}[t]
\includegraphics[scale=0.26]{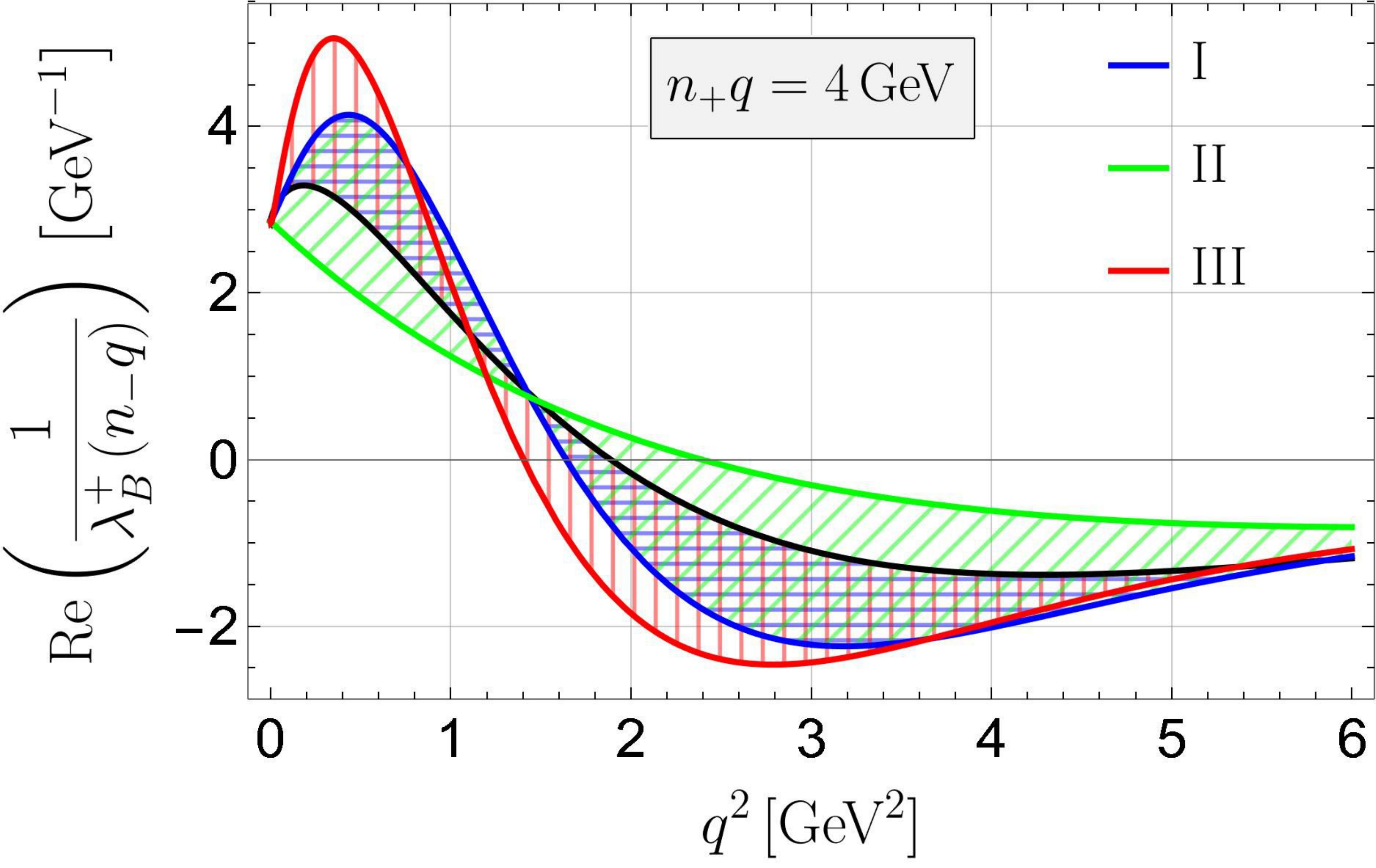}\hspace{0.6cm}
\centering
\includegraphics[scale=0.26]{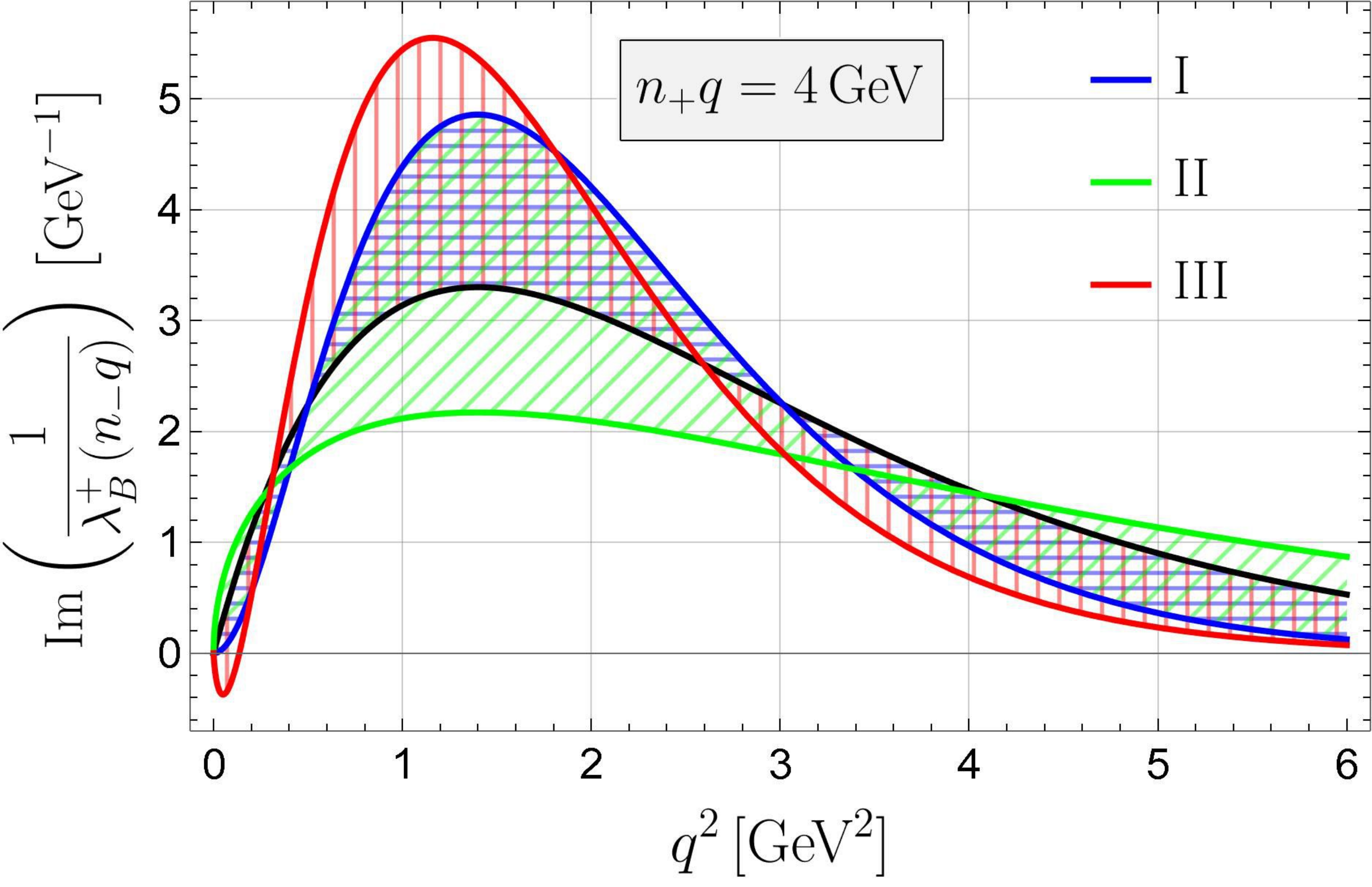}\hskip0.6cm 
\caption{The $q^2$ dependence of real (left) and imaginary (right) parts of $1/\lambda_B^+(n_-q)$ at fixed $\nplusq = 4$ GeV for $\phi^{\rm{I,II, III}}(\omega)$ in blue, green and red, respectively. The bands are obtained by varying the model parameter $a$ within its range indicated in \eqref{eq:models} for 
fixed $\lambda_B=350$ MeV.  }
\label{fig:lamplus}
\end{figure}

\begin{figure}[t]
\vskip0.2cm
\includegraphics[scale=0.25]{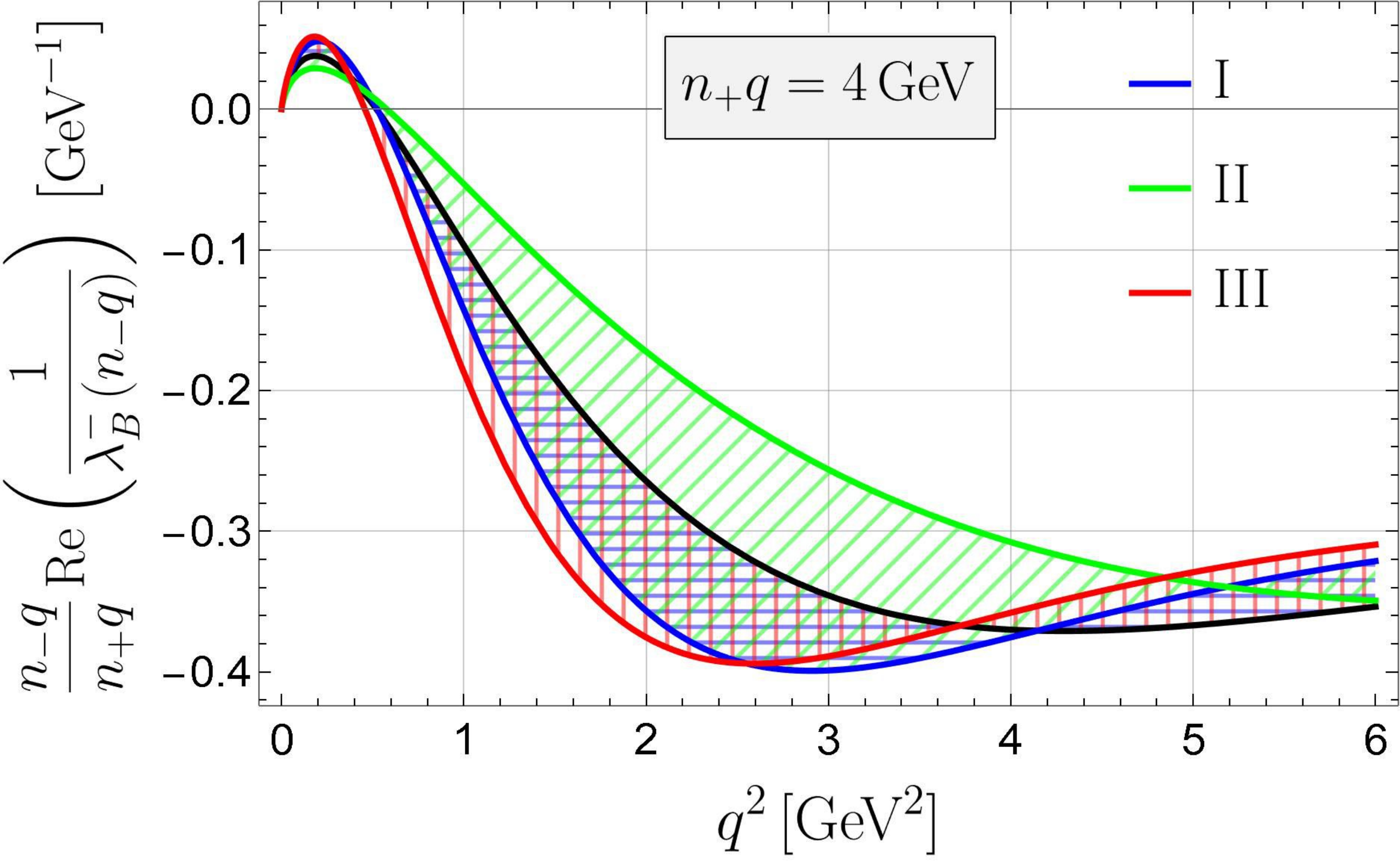}
\hspace{0.6cm}
\centering
\includegraphics[scale=0.25]{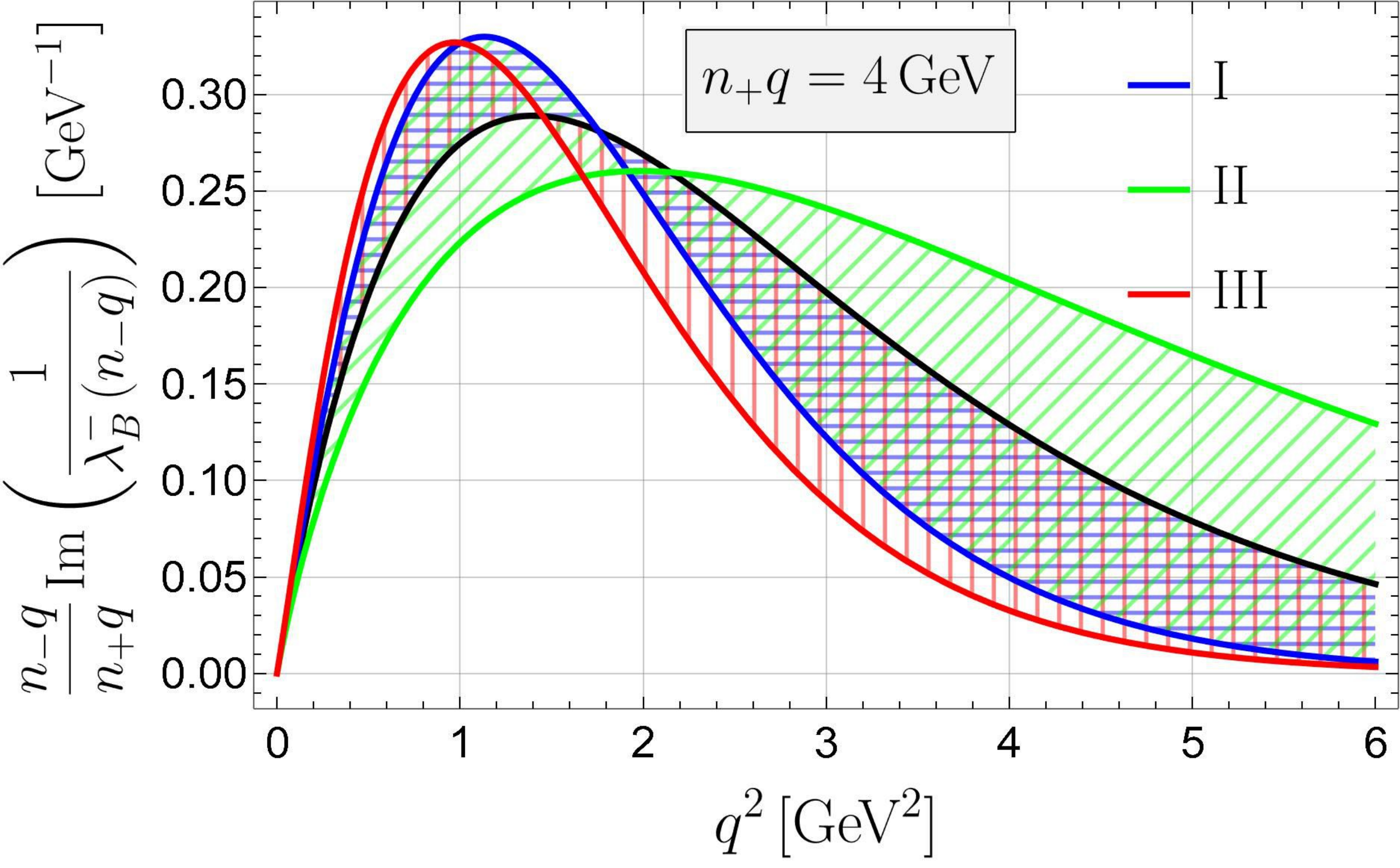}\hskip0.6cm 
\vskip0.2cm
\caption{ As Fig.~\ref{fig:lamplus} but for $\frac{n_-q}{n_+q}\times 1/\lambda_B^-(n_-q)$.}
\label{fig:lammin}
\end{figure}

To study this dependence, we consider three two-parameter models  \cite{Beneke:2018wjp}
\begin{align}
  \phi_+^{\rm I}(\omega)& = \left[(1-a) + \frac{a \omega}{2\omega_0}\right] \frac{\omega}{\omega_0^2} e^{-\omega/\omega_0}  \quad\quad 0\leq a \leq 1 \nonumber \\
       \phi_+^{\rm II}(\omega)& = \frac{1}{\Gamma(2+a)} \frac{\omega^{1+a}}{\omega_0^{2+a}} e^{-\omega/\omega_0} \quad\quad -0.5<a<1 \nonumber \\
\phi_+^{\rm III}(\omega)& = \frac{\sqrt{\pi}}{2 \Gamma(3/2+a)} \frac{\omega}{\omega_0^2} e^{-\omega/\omega_0}U(-a,3/2-a,\omega/\omega_0)\quad\quad 0<a<0.5 \ ,
\label{eq:models}
\end{align}
where $U(\alpha, \beta, z)$ is the confluent hypergeometric function of the second kind. Given $(\omega_0,a)$ one determines $\lambda_B$ and the dimensionless shape parameter $\widehat{\sigma}_1$, related to the first inverse-logarithmic moment. The range of $a$ is chosen such that the range 
$-0.693147 < \widehat{\sigma}_1 <0.693147$ is covered, 
where $\widehat{\sigma}_1 =0$ in the exponential model, 
see \cite{Beneke:2018wjp} for more details. To study the 
influence of the shape of the LCDA, we are then 
interested in the envelope of theoretical predictions 
of all three models spanned by the variation of $a$ 
for given $\lambda_B$. For simplicity, we assume that 
these forms of the $B$ meson LCDA hold at the scale 
$\mu_{hc}=1.5~$GeV, so that no renormalization group 
evolution to the hard-collinear scale needs to be performed.
We obtain $\phi_-(\omega)$ using the Wandzura-Wilczek (WW) relation \cite{Beneke:2000wa}
\begin{equation}
    \phi_-(\omega) = \int_\omega^\infty \frac{d\omega'}{\omega'} \phi_+(\omega') \ .
\end{equation}
The $n_-q$ dependent moments $\lambda_B^\pm(\nm q)$ are then obtained using \eqref{eq:q2moments} (and equivalently for $\lambda_B^-(\nm q)$). Again we define $\lambda_B \equiv \lambda^+_B(n_-q=0)$, such that $\omega_0$ can be related to $\lambda_B$ via 
\begin{align}
\label{eq:omegarel}
    \omega_0^{\rm I} & = \lambda_B \left(1-\frac{a}{2}\right) \ ,\nonumber \\
       \omega_0^{\rm II} & = \frac{\lambda_B}{1+a}  \ , \nonumber\\
          \omega_0^{\rm III} & = \frac{\lambda_B}{1+2 a} \ .
    \end{align}
In Figures~\ref{fig:lamplus} and~\ref{fig:lammin}, 
respectively, we show the $q^2$ dependence of 
$1/\lambda_B^+(n_-q)$ and  $1/\lambda_B^-(n_-q)$ for fixed $\nplusq = 4$ GeV for the three $B$ LCDA models by varying the parameter $a$ within the ranges indicated in \eqref{eq:models}, fixing $\lambda_B=350$ MeV. The black solid line represents the exponential model. There is a significant dependence of $1/\lambda_B^\pm(n_-q)$ on the $B$-meson LCDA shape -- this is expected, as for instance, the $q^2$ dependence of 
the imaginary part $1/\lambda_B^\pm(n_-q)$ is directly 
related to the $\omega$-dependence of $\phi_\pm(\omega)$.

\begin{figure}[t]
\includegraphics[scale=0.30]{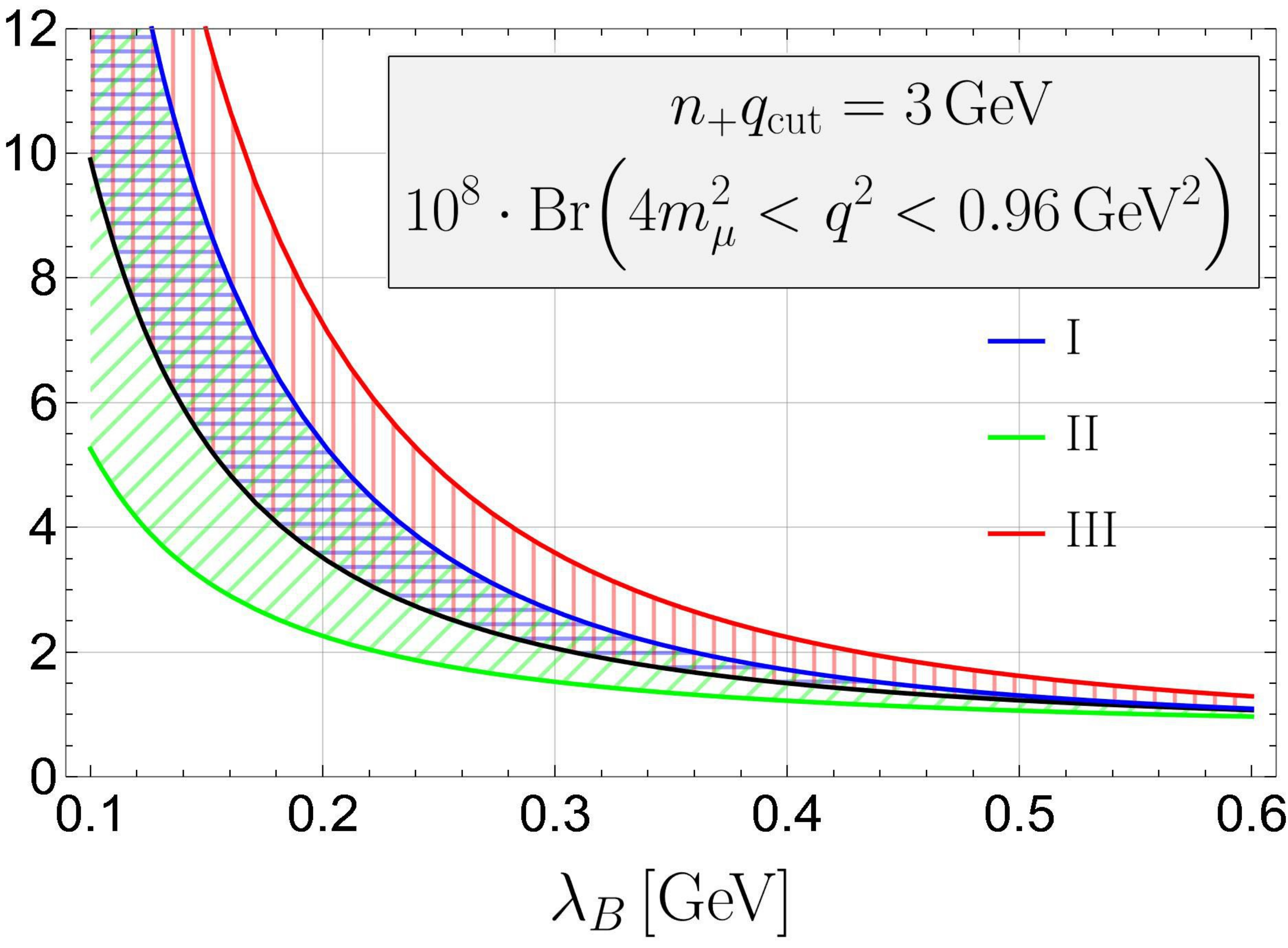}
\centering
\caption{Branching ratio (in $10^{-8}$) of the 
$B^-\to \mu^- \mu^+\,e^- \,\bar{\nu}_{e}$ decay mode in 
the $[4m_\mu^2, 0.96]$~GeV$^2$ $q^2$ bin for the three $B$ LCDA models of \eqref{eq:models} as a function of $\lambda_B$ at 
scale $1.5~$GeV. The bands are obtained by varying the model parameters within their allowed ranges. The solid black curve refers to the exponential model.}
\label{fig:lamdadep}
\end{figure}

Finally, we compute the effect of the $B$ meson LCDA shape on the branching ratio. In Figure~\ref{fig:lamdadep}, we show the dependence of the branching ratio in the  $[4m_\mu^2, 0.96]$~GeV$^2$ $q^2$ bin on $\lambda_B$ and the shape parameter $a$ for the three models. For given  $\lambda_B$ on the horizontal axis, the bands are obtained by varying $a$ in its allowed range. In black, we also show the exponential model. Comparing with our previous results, we observe that the dependence on the shape 
is about as large as the dependence on the $r_{\rm LP}$ variation from the power-suppressed form factor, see Figure~\ref{fig:brlambda}. The conclusion is thus similar to the case of 
 $B\to \gamma \ell \nu$  \cite{Beneke:2018wjp}. Once sufficient data is available, a correlated determination of $\lambda_B$ together with the shape parameter $\widehat{\sigma}_1$ (and, perhaps, others) should be performed. The important point is that the predicted branching fractions are highly sensitive to  $B$ meson LCDA input, even if not necessarily $\lambda_B$ alone. 

\section{Conclusion}

Motivated by the first search and upper limit~\cite{Aaij:2018pka}
for the rare charged-current $B$ decay to a four-lepton final state 
$\ell \bar{\nu}_\ell \ell^{(\prime)} \bar{\ell}^{(\prime)}$, 
this work considered the calculation of the decay amplitude 
with factorization methods. Combining methods previously applied 
to $B^- \rightarrow \lep^- \bv_{\lep}\gamma $ \cite{Beneke:2011nf}
and $B_s\to \mu^+\mu^-\gamma$ \cite{Beneke:2020fot}, we 
obtain the $B\to \gamma^*$ form factors, which depend on the 
invariant masses of the two lepton pairs, in QCD factorization 
at next-to-leading order in $\alpha_s$ and leading power 
in an expansion in $\Lambda_{\rm QCD}/m_b$, and to leading 
order in $\alpha_s$ at next-to-leading power. To this we 
added a simple Breit-Wigner parametrization of the $\rho$, $\omega$ 
intermediate resonances. Although suppressed beyond next-to-leading 
power, the resonances dominate the spectrum in the $\ell^{(\prime)} \bar{\ell}^{(\prime)}$ invariant mass $\sqrt{q^2}$ locally, 
making the 
predictions more uncertain in this region than at large invariant 
mass or for $B^- \rightarrow \lep^- \bv_{\lep}\gamma $. Quite 
generally it must be noted that the parametric counting that 
justifies the heavy-quark expansion is not well respected, 
as is evidenced by the large contribution of the longitudinal 
polarization state of the intermediate virtual photon. 

Our calculations predict branching fractions of a few times 
$10^{-8}$ in the $q^2$ bin up to 1~GeV$^2$, which are accessible 
to the LHC experiments. The branching fraction rapidly drops with 
increasing $q^2$, reaching $10^{-9}$ in the bin 
$[1.5,6]~$GeV$^2$.

Confronting these results to measurements checks our understanding 
of Standard Model dynamics in these rare decays. An important 
further motivation for this investigation has been to explore 
the sensitivity of the decay rate to the inverse moment  
$\lambda_B$ of the leading-twist $B$ meson light-cone 
distribution amplitude. For non-vanishing $q^2$ the access
to $\lambda_B$ is less direct than in 
$B^- \rightarrow \lep^- \bv_{\lep}\gamma$, and requires 
some knowledge of the shape of the LCDA as well. At large 
$q^2$, the sensitivity disappears. We find these expectations 
confirmed in Fig.~\ref{fig:brlambda}, which shows that 
$\lambda_B$ is best determined from the small-$q^2$ bin. 
In this bin the sensitivity to $\lambda_B$ is almost 
comparable to $B^- \rightarrow \lep^- \bv_{\lep}\gamma$ when 
$\lambda_B<200\,$MeV, but less when it is larger. However, 
one should be aware that in this bin the resonance contribution 
is sizeable and there is unquantified model dependence 
related to its interference with the factorization 
contribution. The $q^2$ bin above 1~GeV$^2$ can still yield 
useful bounds on $\lambda_B$, despite its weaker sensitivity. 
As for the case of $B\to \gamma \ell \nu$ \cite{Beneke:2018wjp}, once sufficient data is available, a correlated determination of $\lambda_B$ together with $B$ meson LCDA shape parameters should be performed. Overall, we conclude that the four-lepton final state cannot 
fully replace the $B^- \rightarrow \lep^- \bv_{\lep}\gamma$ mode to 
measure $\lambda_B$. However, given the current state of 
knowledge, any complementary experimental result on $\lambda_B$ 
is worthwhile pursuing.

\subsubsection*{Note added}

While this paper was being finalized, Ref.~\cite{Bharucha:2021zay} 
appeared. We note the following important differences: 
(1) The third independent form factor, related to 
$F_{A_\parallel}$, is missed, see App.~\ref{sec:app:Tdecomp}. 
(2) The $q^2$ distribution is computed without a cut on 
$n_+ q$, hence includes significant phase-space regions 
where the adopted QCD factorization treatment is not applicable.
(3) The residual scale dependence of the leading-power 
form factor at NLO in the strong coupling is much larger 
than ours. Presumably this is because it is assumed (incorrectly)  
that the form of the exponential model is preserved by 
renormalization group evolution.
(4) Only the region of small $q^2<1~$GeV$^2$ is discussed. 
In this region our results are dominated locally by the 
Breit-Wigner parameterization of the $\rho$ and 
$\omega$ resonances, whereas \cite{Bharucha:2021zay} adopts 
the QCD sum rule expression \cite{Braun:2012kp} 
for the power-suppressed form 
factor $\xi$, but in the time-like region.
(5) For the case of identical lepton flavours, we present 
partially integrated branching fractions that correspond 
to experimental observables.

\subsubsection*{Acknowledgements}
We thank A.~Khodjamirian, D.~van Dyk and R.~Zwicky for discussions. 
This research was funded in part by the Deutsche 
Forschungsgemeinschaft (DFG, German Research Foundation) -- 
Project-ID 196253076 -- TRR 110 ``Symmetries and the Emergence of 
Structure in QCD''.

\appendix

\section{Decomposition of the hadronic tensor}
\label{sec:app:Tdecomp}

Using Lorentz covariance only, the most general decomposition of the hadronic tensor $T^{\mu \nu}(p,q)$ defined in~\eqref{eq:hadtensor} contains six independent scalar form factors $F_{1\dots 6} = F_{1\dots 6}(k^2,q^2)$:
\begin{align}
\label{eq:generalTdecom}
    T^{\mu \nu} = F_1 g^{\mu\nu} + F_2 \eps^{\mu\nu\alpha\beta} k_\alpha q_\beta + F_3 k^\mu q^\nu + F_4 q^\mu k^\nu + F_5 k^\mu k^\nu + F_6 q^\mu q^\nu \,.
\end{align}
For $\gamma^* \to \bar{\ell}' \ell'$ and $W^* \to \ell \bar{\nu}_\ell$, the $q^\mu$ ($k^\nu$) terms do not contribute to the decay amplitude if $\ell'$ ($\ell$) is massless.
The Ward identity $q_\mu T^{\mu\nu} = f_B (k+q)^\nu$ implies the relations
\begin{align}
 F_1 + F_3 q\cdot k + F_6 q^2 = f_B \qquad \text{and} \qquad F_4 q^2 + F_5 q\cdot k = f_B \,,    
\end{align}
and hence reduces the number of independent form factors to four.
We choose to eliminate $F_3$ and $F_5$, and write
\begin{align}
\label{eq:TafterWard}
    T^{\mu\nu} = \, &F_1 g^{\mu\nu} + F_2 \eps^{\mu\nu\alpha\beta} k_\alpha q_\beta + \left( \frac{f_B - F_1 - q^2 F_6}{k\cdot q}\right) k^\mu q^\nu \nonumber \\ 
    &+ F_4 q^\mu k^\nu + \left( \frac{f_B - q^2 F_4}{k\cdot q}\right) k^\mu k^\nu + F_6 q^\mu q^\nu \,.
\end{align}
Since we consider massless leptons we now can drop all terms in the second line, which leaves three independent form factors.
The number of independent form factors can be associated with the number of independent polarization states of the virtual photon.
Note that dropping $F_{4,5,6}$ in~\eqref{eq:generalTdecom} \textit{before} applying the Ward identity would lead to the omission of the 
$F_6$ term in the coefficient of the $k^\mu q^\nu$ term
 in~\eqref{eq:TafterWard}, and to the wrong  conclusion (since $F_6$ 
does not vanish) that there are 
only two independent form factors $F_{1,2}$ for massless 
leptons.\footnote{Since $F_6$ is multiplied by $q^2$, this 
conclusion {\em is} correct for the $B\to \gamma$ form factors.}
It is straightforward to work out the relations between the form factors $F_{1,2,6}$ and $F_{A_\perp, V, A_\parallel}$ used in the main text.

\section{Angular distribution}

For non-identical leptons, we find that the full five-fold 
differential branching fraction
\label{app:angdist}
\begin{equation}
\frac{d^5 {\rm Br}\left(B^- \rightarrow \lep \, \bv_{\lep} \, \lep' \bar{\lep'}\right)}{dq^2\,dk^2\, d\cos \theta_{\gamma}\, d\cos \theta_{W} \,d \phi}=\frac{\tau_BG_F^2|V_{ub}|^2\alem^2}{2^{12}\pi^4m_B^5}\frac{\sqrt{\lambda}}{q^4}\,\sqrt{1-\frac{4m_{\ell'}^2}{q^2}} \left(1-\frac{m_\ell^2}{k^2} \right)\,f\left(q^2,k^2,{\bf \Omega}\right) \,,
\end{equation}
with ${\bf \Omega} = (\theta_\gamma, \theta_W, \phi)$,
assuming vanishing lepton masses, 
can be expressed in terms of nine independent angular coefficient functions $K_i = K_i(k^2,q^2)$, $i = 1 \dots 9$, as follows:
\begin{align}
f\left(q^2,k^2,{\bf \Omega}\right) &= K_{1} \sin^2 \theta_W \sin^2 \theta_\gamma + K_{2} \left(1+\cos^2\theta_W\right)\left(1+\cos^2\theta_\gamma\right) + K_{3} \sin^2\theta_W \sin^2\theta_\gamma \sin^2\phi \nonumber \\
&+K_{4}\cos \theta_W \left(1 + \cos^2\theta_\gamma\right) \nonumber \\
&+\Big(K_{5}+ K_{6}\cos \theta_W \Big) \sin \theta_W \sin \theta_\gamma \cos \theta_\gamma \sin \phi \nonumber\\
&+\Big(K_{7}+ K_{8}\cos \theta_W \Big) \sin \theta_W \sin \theta_\gamma \cos \theta_\gamma \cos \phi\nonumber\\
&+K_{9} \sin^2 \theta_W \sin^2 \theta_\gamma \cos  \phi \sin \phi \ .
\end{align}
Our definition of the helicity angles corresponds to the one in~\cite{Boer:2014kda} (with the replacements $\theta_\Lambda^{\text{\cite{Boer:2014kda}}} \to \theta_\gamma$, $\theta_\ell^{\text{\cite{Boer:2014kda}}} \to \theta_W$ and $\phi^{\!\text{\cite{Boer:2014kda}}} \to \phi$),
i.e. $\theta_W$ is the angle between $\vec{p}_\ell$ and the $z$ axis in the rest frame of the $\ell \bar{\nu}_\ell$ system, $\theta_\gamma$ is the angle between $\vec{q}_1$ and the $z$ axis in the dilepton rest frame, and $\phi$ is the relative angle between the decay planes.
Introducing $f_{A_\parallel} = \lambda \Fn$, $f_V = 2\sqrt{k^2 q^2 \lambda} F_V$, and $f_{A_\perp} = 2\sqrt{k^2 q^2} (m_B^2 - k^2 + q^2) \Fa$ uniquely determines the kinematic functions that multiply the form factors, and we find:
\begin{align}
    K_{1} &= \frac14 \left(|f_{A_\perp}|^2 - |f_V|^2 + 2 |f_{A_\parallel}|^2\right) \\
    K_{2} &= \frac14 \left( |f_{A_\perp}|^2 + |f_V|^2 \right) \\
    K_{3} &= -\frac12 \left( |f_{A_\perp}|^2 - |f_V|^2 \right) \\[0.2cm]
      K_{9} &= - \operatorname{Im}\left(f_{A_\perp} f_V^*\right) \\
    K_{4} &= -\operatorname{Re}\left(f_{A_\perp} f_V^*\right)\\ 
  K_{6} &= -\operatorname{Im}\left(f_{A_\parallel} f_V^*\right) \\
  K_{7} &= -\operatorname{Re}\left(f_{A_\parallel} f_V^*\right)\\
         K_{5} &= \operatorname{Im}\left(f_{A_\parallel} f_{A_\perp}^*\right) \\
  K_{8} &= \operatorname{Re}\left(f_{A_\parallel} f_{A_\perp}^*\right) \,.
\end{align}



\bibliography{refs.bib}

\end{document}